\documentclass[draft,tightenlines,nofootinbib,preprint,aps,eqsecnum]{revtex4}
\begin{document}

\title{Non-Inertial Frames in Minkowski Space-Time, Accelerated either
Mathematical or Dynamical Observers and Comments on Non-Inertial
Relativistic Quantum Mechanics}

\author{Horace W. Crater}
\affiliation{The University of Tennessee Space Institute \\
Tullahoma, TN 37388 USA \\
E-mail: hcrater@utsi.edu}
\author{Luca Lusanna}
\affiliation{Sezione INFN di Firenze\\
Polo Scientifico\\
Via Sansone 1\\
50019 Sesto Fiorentino (FI), Italy\\
E-mail: lusanna@fi.infn.it}

\begin{abstract}
After a review of the existing theory of non-inertial frames and
mathematical observers in Minkowski space-time we give the explicit
expression of a family of such frames obtained from the inertial ones by
means of point-dependent Lorentz transformations as suggested by the
locality principle. These non-inertial frames have non-Euclidean 3-spaces
and contain the differentially rotating ones in Euclidean 3-spaces as a
subcase.

Then we discuss how to replace mathematical accelerated observers with
dynamical ones (their world-lines belong to interacting particles in an
isolated system) and of how to define Unruh-DeWitt detectors without using
mathematical Rindler uniformly accelerated observers. Also some comments are
done on the transition from relativistic classical mechanics to relativistic
quantum mechanics in non-inertial frames.
\end{abstract}

\today

\maketitle

\vfill\eject

\section{Introduction}

The theory of non-inertial frames in special relativity (SR) is a topic
rarely discussed and till recently there was no attempt to develop a
consistent general theory of it. All the results of the standard model of
elementary particles are defined in the inertial frames of Minkowski
space-time. Only at the level of nuclear, atomic and
molecular physics one needs a local study of non-inertial frames in SR, for
instance the rotating ones for the Sagnac effect.

\medskip

As can be seen in the review paper \cite{1} most of the papers treating
accelerated observers are concentrated on the Rindler uniformly accelerated
ones due to their relevance for the Unruh-DeWitt effect. However these
observers form a peculiar set existing only due to the Lorentz signature of
Minkowski space-time: since they are asymptotic to the light-cone at past
and future time infinity, they disappear in the non-relativistic (NR) limit
together with the light-cone and do not identify any accelerated observer in
Galilei space-time.

\medskip

A first consistent theory of global non-inertial frames and accelerated
observers in Minkowski space-time, together with their limit to Galilei
space-time, was developed in the papers of Refs.\cite{2}. It was motivated by relativistic
metrology \footnote{
See Ref.\cite{3} for an updated review on relativistic metrology on Earth
and in the Solar System.} with its problem of clock synchronization, by the
problem of the elimination of the relative times in relativistic bound
states (absence of time-like excitations in spectroscopy) so as to arrive at
a consistent formulation of relativistic quantum mechanics (RQM) \cite{4}
and by the necessity to have a formulation of non-inertial frames extendible
to general relativity (GR) at least in the Post-Newtonian approximation used
by space physics around the Earth and by astronomical conventions in the
Solar System and outside it (see Ref.\cite{3}). See Ref.\cite{5} for a
review on the use of this theory of non-inertial frames in special (SR) and
general (GR) relativity.

\bigskip

The description of non-inertial frames in SR is highly non trivial because,
due to the Lorentz signature of Minkowski space-time, time is no longer
absolute and there is no notion of instantaneous 3-space: the only intrinsic
structure is the conformal one, i.e. the light-cone as the locus of incoming
and outgoing radiation. \textit{A convention on the synchronization of
clocks is needed to define an instantaneous 3-space}. For instance the
\textit{1-way velocity of light} from one observer A to an observer B has a
meaning only after a choice of a convention for synchronizing the clock in A
with the one in B. Therefore the crucial quantity in SR is the \textit{2-way
(or round trip) velocity of light $c$} involving only one clock. It is this
velocity which is isotropic and constant in SR and replaces the standard of
length in relativistic metrology \cite{3}. \medskip

The Einstein convention for the synchronization of clocks in Minkowski
space-time uses the 2-way velocity of light to identify the Euclidean
3-spaces of the inertial frames centered on an inertial observer A by means
of only his/her clock. The inertial observer A sends a ray of light at $
x_{i}^{o} $ towards the (in general accelerated) observer B; the ray is
reflected towards A at a point P of B's world-line and then reabsorbed by A
at $x_{f}^{o}$; by convention P is synchronous with the mid-point between
emission and absorption on A's world-line, i.e. $x_{P}^{o}=x_{i}^{o}+{\frac{1%
}{2}}\,(x_{f}^{o}-x_{i}^{o})={\frac{1}{2}}\,(x_{i}^{o}+x_{f}^{o})$. This
convention selects the Euclidean instantaneous 3-spaces $x^{o}=ct=const.$ of
the inertial frames centered on A. Only in this case does the one-way
velocity of light between A and B coincides with the two-way one, $c$.
However if the observer A is accelerated, the convention can break down due
to the possible appearance of coordinate singularities.

\bigskip

As reviewed in Ref.\cite{2}, the existing coordinatizations centered on
accelerated observers, like either Fermi or Riemann-normal coordinates, hold
only locally They are based on the \textit{1+3 point of view} \cite{6}, in
which only the world-line of a time-like observer is given. In each point of
the world-line the observer 4-velocity determines an orthogonal
3-dimensional space-like tangent hyper-plane, which is identified with an
instantaneous 3-space. However, these tangent planes intersect at a certain
distance from the world-line (the so-called acceleration length depending
upon the 4-acceleration of the observer \cite{7}), where 4-coordinates of
the Fermi type develop a coordinate singularity. Another type of coordinate
singularity is developed in rigidly rotating coordinate systems at a
distance $r$ from the rotation axis where $\omega\, r = c$ ($\omega$ is the
angular velocity and $c$ the two-way velocity of light). This is the
so-called "horizon problem of the rotating disk": a time-like 4-velocity
becomes a null vector at $\omega\, r = c$, like it happens on the horizon of
a black-hole.

\medskip

As a consequence, a theory of global non-inertial frames in Minkowski
space-time has to be developed in a metrology-oriented way to overcame the
pathologies of the 1+3 point of view. This has been done in the papers of
Ref.\cite{2} by using the \textit{3+1 point of view} in which, besides the
world-line of a time-like observer, one gives a global nice foliation of the
space-time with instantaneous 3-spaces. In Section II we give a review of
the status of the theory.

\bigskip

In Ref.\cite{2} there is the list of conditions needed to avoid every kind
of pathology in the definition of non-inertial frames. Since they are
complicated non-linear restrictions, till now only differentially rotating
non-inertial frames in Euclidean 3-spaces are completely under control.

\medskip

The aim of this paper is to extend this class to a family of non-inertial
frames with non-Euclidean 3-spaces obtainable from inertial frames by means
of point-dependent Lorentz transformations as suggested by the locality
principle \cite{7} (at each instant an accelerated detector gives the same
data of an instantaneously comoving inertial detector). This will be done in
Section III and in Appendix A.

\medskip

Then in Section IV we will make some comments on the nature (either
mathematical or dynamical) of the observers on which the non-inertial frames
are centered. In particular we will study how to compare the descriptions
given two mathematical observers (Alice and Bob), starting from the case in
which one of them is the origin of the inertial rest frame of an isolated
system. We will also show that at the classical level it is possible to have
dynamical observers by using the world-lines of dynamical particles
contained in the isolated system as origin of the non-inertial frame.

\medskip

In Section V we will make some comments on how to extend  RQM
and relativistic entanglement (in the formulation of Refs. \cite{4,8}) to
non-inertial frames, on what could be the meaning of a "quantum observer"
and on the description of Unruh-DeWitt detectors \cite{1} in our framework.
\medskip

Finally in the Conclusions we will delineate some open problems.

\section{ Review on Non-Inertial Frames}

After a review of global non-inertial frames and the description of isolated
systems in them by means of parametrized Minkowski theories (Subsections A
and B), we introduce the inertial rest frame of isolated systems, whose
relativistic collective variables and  Wigner covariant 3-variables in
the rest Wigner 3-spaces are then given (Subsections C and D). In Subsection
E there is the expression of differentially rotating frames, while in
Subsection F there is the form of the rest-frame conditions in the
non-inertial rest frames of isolated systems. The isolated systems described
here consist only of scalar massive positive-energy particles, because in
Section IV we will use one of them as a dynamical observer origin of the
non-inertial frame.

\subsection{Global Non-Inertial Frames from the 3+1 Point of View}

Assume that the world-line $x^{\mu}(\tau)$ of an arbitrary time-like
observer \footnote{
An observer, or better a mathematical observer, is an idealization of a
measuring apparatus containing an atomic clock and defining, by means of
gyroscopes, a set of spatial axes (and then a, maybe orthonormal, tetrad
with a convention for its transport) in each point of the world-line.}
carrying a standard atomic clock is given: $\tau$ is an arbitrary
monotonically increasing function of the proper time of this clock. Then one
gives an admissible 3+1 splitting of Minkowski space-time, namely a nice
foliation with space-like instantaneous 3-spaces $\Sigma_{\tau}$. It is the
mathematical idealization of a protocol for clock synchronization: all the
clocks in the points of $\Sigma_{\tau}$ sign the same time of the atomic
clock of the observer \footnote{
Actually the physical protocols (think of GPS) can establish a clock
synchronization convention only inside future light-cone of the physical
observer defining the local 3-spaces only inside it.}. The observer and the
foliation define a global non-inertial reference frame after a choice of
4-coordinates. On each 3-space $\Sigma_{\tau}$ one chooses curvilinear
3-coordinates $\sigma^r$ having the observer as origin. The quantities $
\sigma^A = (\tau; \sigma^r)$ are the Lorentz-scalar and observer-dependent
\textit{radar 4-coordinates}, first introduced by Bondi \cite{9}. \medskip

Giving the whole world-line of an arbitrary time-like observer and moreover
a nice foliation with 3-spaces is a \textit{non-factual} necessity required
by the Cauchy problem. Once we have given the Cauchy data on the initial
Cauchy surface (a un-physical process), we can predict the future with every
observer receiving the information only from his/her past light-cone
(retarded information from inside it; electromagnetic signals on it)
\footnote{%
As far as we know the theorem on the existence and unicity of solutions has
not yet been extended starting from data given only on the past light-cone.}%
.. For non-relativistic observers the situation is simpler, but the
non-factual need of giving the Cauchy data on a whole initial absolute
Euclidean 3-space is present also in this case for non-relativistic field
equations like the Euler equation for fluids.

\bigskip

If $x^{\mu} \mapsto \sigma^A(x)$ is the coordinate transformation from the
Cartesian 4-coordinates $x^{\mu}$ of an inertial frame centered on a \textit{%
reference inertial observer} to radar coordinates, its inverse $\sigma^A
\mapsto x^{\mu} = z^{\mu}(\tau ,\sigma^r)$ defines the \textit{embedding}
functions $z^{\mu}(\tau ,\sigma^r)$ describing the 3-spaces $\Sigma_{\tau}$
as embedded 3-manifolds into Minkowski space-time. The induced 4-metric on $
\Sigma_{\tau}$ is the following functional of the embedding: ${}^4g_{AB}(\tau
,\sigma^r) = [z^{\mu}_A\, \eta_{\mu\nu}\, z^{\nu}_B](\tau ,\sigma^r)$, where
$z^{\mu}_A = \partial\, z^{\mu}/\partial\, \sigma^A$ and ${}^4\eta_{\mu\nu}
= \epsilon\, (+---)$ is the flat metric \footnote{$\epsilon = \pm 1$
according to either the particle physics $\epsilon = 1$ or the general
relativity $\epsilon = - 1$ convention.}.

\medskip

While the 4-vectors $
z^{\mu}_r(\tau ,\sigma^u)$ are tangent to $\Sigma_{\tau}$, so that the unit
normal $l^{\mu}(\tau ,\sigma^u)$ is proportional to $\epsilon^{\mu}{}_{
\alpha \beta\gamma}\, [z^{\alpha}_1\, z^{\beta}_2\, z^{\gamma}_3](\tau
,\sigma^u)$, one has $z^{\mu}_{\tau}(\tau ,\sigma^r) = [N\, l^{\mu} + N^r\,
z^{\mu}_r](\tau ,\sigma^r)$ with $N(\tau ,\sigma^r) = \epsilon\,
[z^{\mu}_{\tau}\, l_{\mu}](\tau ,\sigma^r) = 1 + n(\tau, \sigma^r)$ and $%
N_r(\tau ,\sigma^r) = - \epsilon\, {}^4g_{\tau r}(\tau ,\sigma^r)$ being the
lapse and shift functions respectively. The unit normal $l^{\mu}(\tau,
\sigma^u)$ and the space-like 4-vectors $z^{\mu}_r(\tau, \sigma^u)$ identify
a (in general non-ortho-normal) tetrad in each point of Minkowski
space-time. The tetrad in the origin $\Big($ $l^{\mu}(\tau, 0)$ (in general non
parallel to the observer 4-velocity), $z_r^{\mu}(\tau, 0)$ $\Big)$ is a set of axes
carried by the observer; their $\tau$-dependence implies a convention of
transport along the world-line \footnote{
In the 1+3 point of view usually the tetrad carried by the observer has the
unit 4-velocity as time-like vector and often the Fermi-Walker transport of
the tetrad is used.}. See Ref.\cite{5} for the two congruences of time-like
observers (the Eulerian one with 4-velocity field equal to the unit normal and the other,
not surface-forming, with velocity field proportional to $%
z_{\tau}^{\mu}(\tau, \sigma^r)$) associated with each global non-inertial
frame.

\bigskip

Therefore starting from the \textit{four} independent embedding functions $%
z^{\mu}(\tau, \sigma^r)$ one obtains the \textit{ten} components ${}^4g_{AB}$
of the 4-metric, which play the role of the \textit{inertial potentials}
generating the relativistic apparent forces in the non-inertial frame.
For instance the shift functions $N_{r}(\tau ,\sigma ^{u})=-\epsilon \,{}^{4}g(\tau ,\sigma ^{u})$ describe
inertial forces of the gravito-magnetic type induced by the global
non-inertial frame. It can be shown \cite{2} that the usual NR Newtonian inertial potentials are
hidden in these functions. The extrinsic curvature
tensor ${}^3K_{rs}(\tau, \sigma^u) = [{\frac{1}{{2\, N}}}\, (N_{r|s} +
N_{s|r} - \partial_{\tau}\, \epsilon {}^4g_{rs})](\tau, \sigma^u)$, describing the
\textit{shape} of the instantaneous 3-spaces of the non-inertial frame as
embedded 3-sub-manifolds of Minkowski space-time, is a secondary inertial
potential, functional of the ten inertial potentials ${}^4g_{AB}$. \medskip

Now a relativistic positive-energy scalar particle with world-line $%
x_{o}^{\mu }(\tau )$ is described by 3-coordinates $\eta ^{r}(\tau )$
defined by $x_{o}^{\mu }(\tau )=z^{\mu }(\tau ,\eta ^{r}(\tau ))$,
satisfying equations of motion containing relativistic inertial forces with
the correct non-relativistic limit as shown in Ref.\cite{2,10}. Fields have
to be redefined so as to know the clock synchronization convention: for
instance a Klein-Gordon field $\tilde{\phi}(x^{\mu })$ has to be replaced
with $\phi (\tau ,\sigma ^{r})=\tilde{\phi}(z^{\mu }(\tau ,\sigma ^{r}))$.

\bigskip

The foliation is nice and admissible if it satisfies the conditions:
\hfill\medskip

1) $N(\tau ,\sigma^r) > 0$ in every point of $\Sigma_{\tau}$ so that the
3-spaces never intersect, avoiding the coordinate singularity of Fermi
coordinates;\hfill\medskip

2) $\epsilon\, {}^4g_{\tau\tau}(\tau ,\sigma^r) = (N^2 - N_u\, N^u)(\tau, \sigma^r) > 0$, so to avoid the
coordinate singularity of the rotating disk, and with the positive-definite
3-metric $h_{rs}(\tau ,\sigma^u) = - \epsilon\, {}^4g_{rs}(\tau ,\sigma^u)$ ($h = det\, h_{rs}$)
having three positive eigenvalues (these are the M$\o $ller conditions \cite
{11});\hfill\medskip

3) all the 3-spaces $\Sigma_{\tau}$ must tend to the same space-like
hyper-plane at spatial infinity with a unit normal $\epsilon^{\mu}_{\tau}$,
which is the time-like 4-vector of a set of asymptotic ortho-normal tetrads $%
\epsilon^{\mu}_A$. These tetrads are carried by asymptotic inertial
observers and the spatial axes $\epsilon^{\mu}_r$ are identified by the
fixed stars of star catalogues. At spatial infinity the lapse function tends
to $1$ and the shift functions vanish.

\bigskip

By using the asymptotic tetrads $\epsilon^{\mu}_A$ one can give the
following parametrization of the embedding functions

\begin{eqnarray}
z^{\mu}(\tau, \sigma^r) &=& x^{\mu}(\tau) + \epsilon^{\mu}_A\, F^A(\tau,
\sigma^r),\qquad F^A(\tau, 0) = 0,  \nonumber \\
&&{}  \nonumber \\
x^{\mu}(\tau) &=& x^{\mu}_o + \epsilon^{\mu}_A\, f^A(\tau),
 \label{2.2}
\end{eqnarray}

\noindent where $x^{\mu}(\tau)$ is the world-line of the observer. The
functions $f^A(\tau)$ determine the 4-velocity $u^{\mu}(\tau) = {\dot x}%
^{\mu}(\tau)/ \sqrt{\epsilon\, {\dot x}^2(\tau)}$ (${\dot x}^{\mu}(\tau) = {%
\frac{{d x^{\mu}(\tau)}}{{d\tau}}}$) and the 4-acceleration $a^{\mu}(\tau) =
{\frac{{d u^{\mu}(\tau)}}{{d\tau}}}$ of the observer. For an inertial frame
centered on the inertial observer $x^{\mu}(\tau) = x_o^{\mu} + \epsilon^{\mu}_{\tau}\,
\tau$ the embedding is $z^{\mu}(\tau, \sigma^r) = x^{\mu}(\tau) + \epsilon^{\mu}_r\, \sigma^r$,
with $\epsilon^{\mu}_A$ being an ortho-normal tetrad identifying the Cartesian axes. \bigskip

The M$\o $ller conditions are non-linear differential conditions on the
functions $f^{A}(\tau )$ and $F^{A}(\tau ,\sigma ^{r})$, so that it is very
difficult to construct explicit examples of admissible 3+1 splittings.
When these conditions are satisfied Eqs.(2.1)
describe a {\it global non-inertial frame} in Minkowski space-time.

\subsection{Dynamics in Non-Inertial Frames: Parametrized Minkowski Theories
for Isolated Systems}

In this framework one can describe every isolated system (particles, fields,
strings, fluids) admitting a Lagrangian description with \textit{%
parametrized Minkowski theories} \cite{2,12}. One couples the Lagrangian to
an external gravitational field and then replaces the 4-metric with the
4-metric ${}^4g_{AB}(\tau, \sigma^r)$ induced by an admissible foliation.
The new Lagrangian, a functional of the matter described in radar
4-coordinates and of the embedding $z^{\mu}(\tau, \sigma^r)$ (through the
4-metric), allows us to define an action which is invariant under frame
preserving diffeomorphisms \cite{13}. As a consequence, if $T^{\mu\nu}$ is
the energy-momentum tensor of the matter and $\rho_{\mu}(\tau, \sigma^r)$
the canonical momentum conjugate to the embedding, we get the following four
first-class constraints and the following form of the Poincar\'e generators (%
$T_{\perp\perp} = l_{\mu}\, l_{\nu}\, T^{\mu\nu}$, $T_{\perp r} = l_{\mu}\,
z_{r \nu}\, T^{\mu\nu}$, $T_{rs} = z_{r \mu}\, z_{s \nu}\, T^{\mu\nu}$; $%
h^{ru}\, g_{us} = \delta^r_s$ )

\begin{eqnarray}
\mathcal{H}_{\mu}(\tau, \sigma^r) &=& \rho_{\mu}(\tau, \sigma^u) - \sqrt{
h(\tau, \sigma^u)}\, \Big[l_{\mu}\, T_{\perp\perp} - z_{r\mu}\,
h^{rs}\, T_{\perp s}\Big](\tau, \sigma^u) \approx 0,  \nonumber \\
{}&& \{ \mathcal{H}_{\mu}(\tau, \sigma^r_1), \mathcal{H}_{\nu}(\tau,
\sigma^r_2) \} = 0,  \nonumber \\
{}&&  \nonumber \\
P^{\mu } &=& \int d^{3}\sigma \rho ^{\mu }(\tau ,\sigma^u), \qquad J^{\mu
\nu } = \int d^{3}\sigma (z^{\mu }\rho ^{\nu }-z^{\nu }\rho ^{\mu })(\tau
,\sigma^u).  \label{2.3}
\end{eqnarray}

\noindent These constraints imply that the transition among different
non-inertial frames is described as a \textit{gauge transformation} (so that
only the appearances of phenomena change, not the physics) \cite{2,4,5,12,14}%
.. The canonical Hamiltonian is zero and the Dirac Hamiltonian is $H_D = \int
d^3\sigma\, \lambda^{\mu}(\tau, \sigma^r)\, \mathcal{H}_{\mu}(\tau,
\sigma^r) + S_M$ with $\lambda^{\mu}(\tau, \sigma^r)$ arbitrary Dirac
multipliers and $S_M$ a surface term at spatial infinity needed to define
the functional derivatives so that the variation $\delta\, H_D$ is
proportional to the Hamilton equations. This term is the analogue of the
DeWitt surface term in canonical ADM GR: as shown in Ref.\cite{5} in GR this
term is the strong ADM energy, which is equal to the weak ADM energy, i.e.
to a volume integral over the 3-space of the energy density, modulo the
first.class constraints of GR. Here we have $S_M = Mc + constraints (2.2)$
with $Mc = \sqrt{\epsilon\, P^2}$ the mass of the isolated system.

\subsection{The Inertial Rest Frame of Isolated Systems, Their Relativistic
Collective Variables, the Inertial Wigner Rest 3-Space and the External Poincare' Generators}

Inertial frames with Euclidean 3-spaces are a special case of this theory.
For isolated systems there is a special family of inertial systems, the
\textit{intrinsic inertial rest frames}, in which the space-like 3-spaces
are orthonormal to the conserved time-like 4-momentum of the isolated system
\footnote{
The \textit{non-inertial rest frames} are a special class of non-inertial
frames, in which the space-like hyper-planes at spatial infinity are
orthogonal to the conserved 4-momentum of the isolated system. As shown in
Ref.\cite{5} they are important in GR in asymptotically Minkowskian
space-times without super-translations.}.

The internal rest 3-space, named {\it Wigner 3-space}, is defined in such a way
that it is the same for all the reference inertial systems describing it
(modulo a Wigner rotation) and its 3-vectors are Wigner spin-1 3-vectors, so
that the covariance under Poincar\'{e} transformations is under control.

As a consequence it turns out \cite{2,4,5} that at the Hamiltonian level
every isolated system can be described by a decoupled canonical
non-covariant relativistic center of mass (whose spatial part is the
classical counterpart of the Newton-Wigner position operator) carrying a
pole-dipole structure, namely an internal 3-space with a well defined total
invariant mass $M$ and a total rest spin $\vec{S}$ and a well defined
realization of the Poincar\'{e} algebra (\textit{the external Poincar\'{e}
group} for a free point particle, i.e. for the \textit{\ external center of
mass}, whose mass $M$ and spin $\vec S$ are Casimir invariants describing
the matter of the isolated system in a global way).

\medskip

The canonical non-covariant (a pseudo 4-vector) relativistic center of mass $
{\tilde{x}}^{\mu }(\tau )$, the non-canonical covariant (a 4-vector)
Fokker-Pryce center of inertia $Y^{\mu }(\tau )$ and the non-canonical
non-covariant (a pseudo 4-vector) M$\o $ller center of energy $R^{\mu }(\tau
)$ are the only three relativistic collective variables which can be built
only in terms of the Poincar\'{e} generators of the isolated system \cite
{4,15} \footnote{Since the Poincare' generators are integrals of the components of the energy-momentum tensor of the isolated system over the whole rest 3-space, these three collective variables are \textit{non-local} quantities which
cannot be determined with local measurements.}
so that they depend only on the system and nothing external to it.
All of them have the same constant 4-velocity $h^{\mu }=P^{\mu }/Mc$ and
collapse onto the Newton center of mass of the system in the
non-relativistic limit. As shown in Ref. \cite{14} these three variables can
be expressed as known functions of the Lorentz scalar rest time $\tau $, of
canonically conjugate Jacobi data (frozen (fixed $\tau =0)$ Cauchy data) $%
\vec{z}=Mc\,{\vec{x}}_{NW}(0)$, $\vec{h}=\vec{P}/Mc$, (${\vec{x}}_{NW}(\tau
)={\vec{\tilde{x}}}(\tau )$ is the standard Newton-Wigner 3-position; $%
P^{\mu }$ is the external 4-momentum) \footnote{
The use of $\vec{z}$ avoids taking into account the mass spectrum of the
isolated system at the quantum kinematical level and allows one to avoid the
Hegerfeldt theorem (the instantaneous spreading of wave packets with
violation of relativistic causality) in the relativistic quantum mechanics
(RQM) developed in Ref.\cite{4} using this formalism.}, and of the invariant
mass $M$ and rest spin $\vec{S}$. The external Poincar\'{e} generators are
then expressed in terms of these variables. \medskip

\bigskip

The rest frame embedding has the following definition \cite{4,14}

\begin{eqnarray}
z_W^{\mu}(\tau, \vec \sigma) &=& Y^{\mu}(\tau) + \epsilon^{\mu}_r(\vec h)\,
\sigma^r,\qquad \epsilon^{\mu}_r(\vec h) = \Big( h_r; \delta^i_r + {\frac{{%
h^i\, h_r}}{{1 + \sqrt{1 + {\vec h}^2}}}}\Big),  \nonumber \\
&&{}  \nonumber \\
Y^{\mu}(\tau) &=& \Big(\sqrt{1 + {\vec h}^2}\, (\tau + {\frac{{\vec h \cdot
\vec z}}{{Mc}}}); {\frac{{\vec z}}{{Mc}}} + (\tau + {\frac{{\vec h \cdot
\vec z}}{{Mc}}})\, \vec h + {\frac{{\vec S \times \vec h}}{{Mc\, (1 + \sqrt{%
1 + {\vec h}^2})}}} \Big),  \nonumber \\
{\tilde x}^{\mu}(\tau) &=& Y^{\mu}(\tau ) + \Big(0; {\frac{{- \vec S \times
\vec h}}{{Mc\, (1 + \sqrt{1 + {\vec h}^2})}}}\Big),  \label{2.4}
\end{eqnarray}

\noindent where $Y^{\mu}(\tau)$ is the Fokker-Pryce center of inertia and ${
\tilde x}^{\mu}(\tau)$ is the canonical center of mass of the isolated system. For the inertial rest
frame the asymptotic tetrads are $\epsilon^{\mu}_A(\vec h)$ with $%
\epsilon^{\mu}_{\tau}(\vec h) = h^{\mu}$ and with $\epsilon^{\mu}_r(\vec h)$
of Eq.(2.3) \footnote{
They are the columns of the standard Wigner boost for time-like orbits: this
is the source of the Wigner covariance of the Wigner 3-spaces.}. The external
Poincar\'e group has the generators

\begin{eqnarray}
&&P^{\mu} = M c\, h^{\mu} = M\, c\, \Big(\sqrt{1 + {\ \vec h}^2}; \vec h\Big)%
,  \nonumber \\
&&J^{ij} = z^i\, h^j - z^j\, h^i + \epsilon^{ijk}\, S^k,\qquad K^i = J^{oi}
= - \sqrt{1 + {\vec h}^2}\, z^i + {\frac{{(\vec S \times \vec h)^i}}{{1 +
\sqrt{1 + {\vec h}^2}}}},  \nonumber \\
&&{}  \label{2.5}
\end{eqnarray}

\noindent as a consequence of Eqs.(\ref{2.3}). The last term in the boost is
responsible for the Wigner covariance of the 3-vectors in the rest Wigner
3-space $\tau = const.$.

\subsection{The  Relative 3-Variables of Isolated Systems of Relativistic Particles in the
Inertial Rest Frame and the Internal Poincare' Generators}

As already said, the particles of an isolated system are identified by
Wigner-covariant 3-vectors $\eta^r_i(\tau)$. The world-lines of the
particles (and their 4-momenta) are derived notions, which can be rebuilt
given the 3-coordinates, the time-like observer and the axes of the inertial
rest frame \cite{16}.

Let us consider the simple two-particle system of Ref.\cite{16}. The
Wigner-covariant 3-positions and 3-momenta inside the rest Wigner 3-space
are ${\vec \eta}_i(\tau)$, ${\vec \kappa}_i(\tau)$, $i = 1,2$. The
world-lines and the 4-momenta of the two particles are ($V$ is an arbitrary
action-at-a-distance potential)

\begin{eqnarray}
x^{\mu}_i(\tau) &=& z^{\mu}_W(\tau, {\vec \eta}_i(\tau)) = Y^{\mu}(\tau) +
\epsilon^{\mu}_r(\tau)\, \eta^r_i(\tau),  \nonumber \\
p_i^{\mu}(\tau) &=& h^{\mu}\, \sqrt{m_i^2\, c^2 + {\vec \kappa}_i^2(\tau) +
V(({\vec \eta}_1(\tau) - {\vec \eta}_2(\tau))^2)} - \epsilon_r^{\mu}(\vec
h)\, \kappa_{ir}(\tau),  \nonumber \\
&& \epsilon\, p_i^2 = m_i^2\, c^2 + V(({\vec \eta}_1(\tau) - {\vec \eta}%
_2(\tau))^2).  \label{2.6}
\end{eqnarray}

They are 4-vectors but not canonical like in most of the approaches: there
is a \textit{non-commutative structure} induced by the Lorentz signature of
Minkowski space-time \cite{4,16}.

\medskip

In the Wigner 3-space there is another realization of the Poincar\'e algebra
(\textit{the internal Poincar\'e group}) built with the rest 3-coordinates
and 3-momenta of the matter of the isolated system starting from its
energy-momentum tensor: the internal energy is the invariant mass $M\, c^2$
(as said $M c$ is the Hamiltonian inside the rest 3-space, because we have $%
H_D = M c + constraints$) and the internal angular momentum is the rest spin
$\vec S$. Since we are in rest frames the internal 3-momentum must vanish.
Moreover, to avoid a double counting of the center of mass, the internal
center of mass, conjugate to the vanishing 3-momentum, has to be eliminated:
this can be done by fixing the value of the internal Poincar\'e boost. If we put it
equal to zero, this implies \cite{2} that the time-like observer has to be
an inertial observer coinciding with the non-canonical 4-vector describing
the Fokker-Pryce center of inertia of the isolated system. Therefore the
internal realization of the Poincar\'e algebra is unfaithful and inside the
Wigner rest 3-spaces the matter is described by \textit{relative}
3-positions and 3-momenta. \medskip

For the two-particle system the conserved internal Poincar\'{e} generators
are ($T^{\mu \nu }=\epsilon _{A}^{\mu }(\vec{h})\,\epsilon _{B}^{\nu }(\vec{h}
)\,T^{AB}$ is the matter energy-momentum tensor)

\begin{eqnarray}
M\, c &=& \int d^3\sigma\, T^{\tau\tau}(\tau, \sigma^u) = \sum_{i=1}^2\,
\sqrt{m_i^2\, c^2 + {\vec \kappa}^2_i(\tau) + V(({\ \vec \eta}_1(\tau) - {%
\vec \eta}_2(\tau))^2)},  \nonumber \\
{\vec {\mathcal{P}}} &=& \Big(\int d^3\sigma\, T^{\tau r}(\tau, \sigma^u)%
\Big) = \sum_{i=1}^2\, {\vec \kappa}_i(\tau) \approx 0,  \nonumber \\
\vec S &=& \Big({\frac{1}{2}}\, \epsilon^{ruv}\, \int d^3\sigma\, \sigma^u\,
T^{v\tau}(\tau, \sigma^s)\Big) = \sum_{i=1}^2\, {\vec \eta}_i(\tau) \times {%
\vec \kappa}_i(\tau),  \nonumber \\
{\vec {\mathcal{K}}} &=& \Big(- \int d^3\sigma\, \vec \sigma\,
T^{\tau\tau}(\tau, \sigma^u)\Big) = - \sum_{i=1}^2\, {\vec \eta}_i(\tau)\,
\sqrt{ m_i^2\, c^2 + {\vec \kappa}_i^2(\tau) + V(({\vec \eta}_1(\tau) - {%
\vec \eta}_2(\tau))^2)} \approx 0.  \nonumber \\
&&{}  \label{2.7}
\end{eqnarray}

The rest-frame conditions ${\vec{\mathcal{P}}}\approx 0$, ${\vec{\mathcal{K}}
}\approx 0$, imply that the 3-variables ${\vec \eta}_i(\tau)$, ${\vec \kappa}%
_i(\tau)$ are un-physical: the physical canonical variables in the rest
3-space are $\vec{\rho}(\tau) = {\vec{\eta}}_{1}(\tau) - {\vec{\eta}}%
_{2}(\tau)$ and $\vec{\pi}(\tau) = {\frac{ m_{2}}{m}}\,{\vec{\kappa}}%
_{1}(\tau) - {\frac{m_{1}}{m}}\,{\vec{\kappa}}_{2}(\tau)$, ($m=m_{1}+m_{2}$%
). \ Using these relative variables and imposing the rest frame condition
gives for the internal center of mass $\vec{\eta}(\tau)$ (conjugate to $%
\mathcal{\vec{P}\approx }0$)

\begin{eqnarray}
\vec{\eta}(\tau) &=&{\frac{{m_{1}\,{\vec{\eta}}_{1}(\tau) + m_{2}\,{\vec{\eta%
}}_{2}(\tau)}}{m}} \approx {\frac{{m_{1}\,\sqrt{m_{2}^{2}\,c^{2} + H(\tau)}
- m_{2}\,\sqrt{ m_{1}^{2}\,c^{2} + H(\tau)}}}{{m\,(\sqrt{m_{1}^{2}\,c^{2} +
H(\tau)} + \sqrt{m_{2}^{2}\,c^{2} + H(\tau) })}}}\,\vec{\rho}(\tau),
\nonumber \\
{} &&  \nonumber \\
&\Downarrow & \qquad H(\tau) = {\vec{\pi}}^{2}(\tau) + V({\vec{\rho}}%
^{2}(\tau)),  \nonumber \\
{} &&  \nonumber \\
M\,c &\approx &\sqrt{m_{1}^{2}\,c^{2} + H(\tau)} + \sqrt{m_{2}^{2}\,c^{2} +
H(\tau)},\qquad \vec{S}\approx \vec{\rho}(\tau) \times \vec{\pi}(\tau),
\nonumber \\
x_{1}^{\mu }(\tau) &\approx &Y^{\mu }(\tau) + \epsilon _{r}^{\mu }(\vec{h})\,%
{\frac{\sqrt{ m_{2}^{2}\,c^{2} + H(\tau)}}{{Mc}}}\,\rho ^{r}(\tau),
\nonumber \\
x_{2}^{\mu }(\tau) &\approx &Y^{\mu }(\tau) - \epsilon _{r}^{\mu }(\vec{h})\,%
{\frac{\sqrt{ m_{1}^{2}\,c^{2} + H(\tau)}}{{Mc}}}\,\rho ^{r}(\tau).
\label{2.8}
\end{eqnarray}
\bigskip

Therefore besides the non-local features of the relativistic collective
variables there is an intrinsic \textit{\ spatial non-separability} (only
internal relative 3-variables for the whole isolated system due to the
elimination of relative times) forbidding the identification of subsystems
at the physical level: as shown in Refs. \cite{4,8}  this fact generates a notion of relativistic
entanglement in RQM very different from the non-relativistic one. \medskip

See Subsection F for the inertial rest frames centered on inertial observers
different from the Fokker-Pryce center of inertia.

\subsection{Well Defined Global Non-Inertial Frames}

Till now \cite{2} the solution of the M$\o $ller conditions given in
Subsection A is known in the following two cases in which the instantaneous
3-spaces are parallel Euclidean space-like hyper-planes not equally spaced
due to a linear acceleration.\medskip

A) \textit{Rigid non-inertial reference frames with translational
acceleration}. An example are the following embeddings

\medskip

\begin{eqnarray}
z^{\mu }(\tau ,\sigma ^{u}) &=&x_{o}^{\mu }+\epsilon _{\tau }^{\mu }\,f(\tau
)+\epsilon _{r}^{\mu }\,\sigma ^{r},  \nonumber \\
&&{}  \nonumber \\
&&{}^{4}g_{\tau \tau }(\tau ,\sigma ^{u})=\epsilon \,\Big({\frac{{df(\tau )}%
}{{d\tau }}}\Big)^{2},\quad {}^{4}g_{\tau r}(\tau ,\sigma
^{u})=0,{}^{4}g_{rs}(\tau ,\sigma ^{u})=-\epsilon \,\delta _{rs}.  \nonumber
\\
&&{}  \label{2.9}
\end{eqnarray}

\medskip

This is a foliation with parallel hyper-planes with normal $l^{\mu} =
\epsilon^{\mu}_{\tau} = const.$ and with the time-like observer $%
x^{\mu}(\tau ) = x^{\mu}_o + \epsilon^{\mu}_{\tau}\, f(\tau )$ as origin of
the 3-coordinates. The hyper-planes have translational acceleration ${\ddot x%
}^{\mu}(\tau ) = \epsilon^{\mu}_{\tau}\, \ddot f(\tau )$, so that they are
not uniformly distributed like in the inertial case $f(\tau ) = \tau$.

\bigskip

B) \textit{Differentially rotating non-inertial frames} without the
coordinate singularity of the rotating disk. The embedding defining these
frames is

\begin{eqnarray*}
z^{\mu }(\tau ,\sigma ^{u}) &=&x^{\mu }(\tau )+\epsilon _{r}^{\mu
}\,R^{r}{}_{s}(\tau ,\sigma )\,\sigma ^{s}\,\rightarrow _{\sigma \rightarrow
\infty }\,x^{\mu }(\tau )+\epsilon _{r}^{\mu }\,\sigma ^{r}, \\
&&{} \\
R^{r}{}_{s}(\tau ,\sigma ) &=&R^{r}{}_{s}(\alpha _{i}(\tau ,\sigma
))=R^{r}{}_{s}(f(\sigma )\,{\tilde{\alpha}}_{i}(\tau )), \\
&&{} \\
&&0<f(\sigma )<{\frac{{A^{2}}}{{\sigma }}},\qquad {\frac{{d\,f(\sigma )}}{{%
d\sigma }}}\not=0\,\,(Moller\,\,conditions),
\end{eqnarray*}

\begin{eqnarray}
z^{\mu}_{\tau}(\tau ,\sigma^u) &=& {\dot x}^{\mu}(\tau ) -
\epsilon^{\mu}_r\, R^r{}_s(\tau ,\sigma )\, \delta^{sw}\, \epsilon_{wuv}\,
\sigma^u\, {\frac{{\Omega^v(\tau ,\sigma )}}{c}},  \nonumber \\
z^{\mu}_r(\tau ,\sigma^u) &=& \epsilon^{\mu}_k\, R^k{}_v(\tau ,\sigma )\, %
\Big(\delta^v_r + \Omega^v_{(r) u}(\tau ,\sigma )\, \sigma^u\Big),
\label{2.10}
\end{eqnarray}

\noindent where $\sigma =|\vec{\sigma}|$ and $R^{r}{}_{s}(\alpha _{i}(\tau
,\sigma ))$ is a rotation matrix satisfying the asymptotic conditions $%
R^{r}{}_{s}(\tau ,\sigma )\,{\rightarrow }_{\sigma \rightarrow \infty
}\delta _{s}^{r}$, $\partial _{A}\,R^{r}{}_{s}(\tau ,\sigma )\,{\rightarrow }%
_{\sigma \rightarrow \infty }\,0$, whose Euler angles have the expression $%
\alpha _{i}(\tau ,\vec{\sigma})=f(\sigma )\,{\tilde{\alpha}}_{i}(\tau )$, $%
i=1,2,3$. The unit normal is $l^{\mu }=\epsilon _{\tau }^{\mu }=const.$ and
the lapse function is $1+n(\tau ,\sigma ^{u})=\epsilon \,\Big(z_{\tau }^{\mu
}\,l_{\mu }\Big)(\tau ,\sigma ^{u})=\epsilon \,\epsilon _{\tau }^{\mu }\,{%
\dot{x}}_{\mu }(\tau )>0$. In Ref.\cite{2} there is an indirect
demonstration that the three eigenvalues of the 3-metric $h_{rs}(\tau
,\sigma ^{u})$ are positive so that the Moller\thinspace \thinspace
conditions are satisfied.

\medskip

In Eq.(2.9) one uses the notations $\Omega _{(r)}(\tau ,\sigma )_{uv}=\Big(
R^{-1}(\tau ,\vec{\sigma})\,\partial _{r}\,R(\tau ,\sigma )\Big)_{uv}$ and $
\epsilon _{uvr}\,{\frac{{\Omega ^{r}(\tau ,\sigma )}}{c}}=\Big(R^{-1}(\tau
,\sigma )\,\partial _{\tau }\,R(\tau ,\sigma )\Big)_{uv}=\Omega _{(\tau
)}(\tau ,\sigma )_{uv}$, with $\Omega ^{r}(\tau ,\sigma )=f(\sigma )\,\Omega
_{(R)}(\tau ,\sigma )$ ${\hat{n}}^{r}(\tau ,\sigma )$ \footnote{${\hat{n}}
^{r}(\tau ,\sigma )$ defines the instantaneous rotation axis and $0<\tilde{
\Omega}(\tau ,\sigma )<2\,max\,\Big({\dot{\tilde{\alpha}}}(\tau ),{\dot{
\tilde{\beta}}}(\tau ),{\dot{\tilde{\gamma}}}(\tau )\Big)$ as shown in Ref.\cite{2}.} being
the angular velocity. The angular velocity vanishes at spatial infinity and
has an upper bound proportional to the minimum of the linear velocity $
v_{l}(\tau )={\dot{x}}_{\mu }\,l^{\mu }$ orthogonal to the space-like
hyper-planes. When the rotation axis is fixed and $\Omega _{(R)}(\tau
,\sigma )=\omega =const.$, a simple choice for the function $f(\sigma )$ is $
f(\sigma )={\frac{1}{{1+{\frac{{\omega ^{2}\,\sigma ^{2}}}{{c^{2}}}}}}}$
\footnote{
Nearly rigid rotating systems, like a rotating disk of radius $\sigma _{o}$,
can be described by using a function $f(\sigma )$ approximating the step
function $\theta (\sigma -\sigma _{o})$.}. To evaluate the non-relativistic
limit for $c\rightarrow \infty $, where $\tau =c\,t$ with $t$ the absolute
Newtonian time, one chooses the gauge function $f(\sigma )={\frac{1}{{1+{%
\frac{{\omega ^{2}\,\sigma ^{2}}}{{c^{2}}}}}}}\,\rightarrow _{c\rightarrow
\infty }\,1-{\frac{{\omega ^{2}\,\sigma ^{2}}}{{c^{2}}}}+O(c^{-4})$. This
implies that the corrections to rigidly-rotating non-inertial frames coming
from M$\o $ller conditions are of order $O(c^{-2})$ and become important at
the distance from the rotation axis where the horizon problem for rigid
rotations appears.

\bigskip

As shown in the first paper in Refs.\cite{2}, \textit{global rigid rotations
are forbidden in relativistic theories}, because, if one uses the embedding $%
z^{\mu}(\tau ,\sigma^u)= x^{\mu}(\tau ) + \epsilon^{\mu}_r\, R^r{}_s(\tau
)\, \sigma^s$ describing a global rigid rotation with angular velocity $%
\Omega^r = \Omega^r(\tau )$, then the resulting $g_{\tau\tau}(\tau
,\sigma^u) $ violates M$\o $ller conditions, because it vanishes at $\sigma
= \sigma_R = {\frac{1}{{\Omega (\tau )}}}\, \Big[\sqrt{{\dot x}^2(\tau ) + [{%
\dot x}_{\mu}(\tau )\, \epsilon^{\mu}_r\, R^r{}_s(\tau )\, (\hat \sigma
\times \hat \Omega (\tau ))^r]^2}$ $- {\dot x}_{\mu}(\tau )\,
\epsilon^{\mu}_r\, R^r{}_s(\tau )\, (\hat \sigma \times \hat \Omega (\tau
))^r \Big]$ ( $\sigma^u = \sigma\, {\hat \sigma}^u$, $\Omega^r = \Omega\, {%
\hat \Omega}^r$, ${\hat \sigma}^2 = {\hat \Omega}^2 = 1$). At this distance
from the rotation axis the tangential rotational velocity becomes equal to
the velocity of light. This is the \textit{horizon problem} of the rotating
disk (the horizon is often named the \textit{light cylinder}). Let us remark
that even if in the existing theory of rotating relativistic stars one uses
differential rotations, notwithstanding that in the study of the
magnetosphere of pulsars often the notion of light cylinder is still used.

\subsection{The Rest-Frame Conditions in Non-Inertial Frames}

In admissible either inertial or non-inertial frames described by the
embedding (2.1) with asymptotic tetrads $\epsilon_A^{\mu}$, we must consider
the Lorentz transformation connecting them to the  tetrads  $\epsilon^{\mu}_A(\vec h)$ of the
rest frame: $\epsilon^{\mu}_A = \Lambda_A{}^B(\vec h)\,
\epsilon^{\mu}_B(\vec h)$. An isolated system is still described as a
non-local non-covariant decoupled external center of mass with Jacobi data $%
\vec z$, $\vec h$, carrying a pole-dipole structure with an invariant mass
and a spin, whose expression has been found in Ref.\cite{2} and is ($%
l^{\mu}(\tau, \sigma^u) =\epsilon^{\mu}_A\, l^A(\tau, \sigma^u)$ is the unit
normal to the 3-space)

\begin{eqnarray}
Mc &\approx& \int d^3\sigma\, \sqrt{h(\tau , \sigma^u)}\, \Big[%
T_{\perp\perp}\, l^A - T_{\perp s}\, h^{sr}\, \partial_r\, F^A\Big](\tau ,
\sigma^u)\,\,\, \Lambda_A{}^{\tau}(\vec h),  \nonumber \\
&&{}  \nonumber \\
S^r &\approx& {\frac{1}{2}}\, \epsilon^{ruv}\, \int d^3\sigma\, \sqrt{h(\tau
, \sigma^u)}\, \Big[F^C(\tau , \sigma^u)\, \Big(T_{\perp\perp}\, l^D -
T_{\perp s}\, h^{sr}\, \partial_r\, F^D\Big)(\tau , \vec \sigma) -  \nonumber
\\
&-& F^D(\tau , \vec \sigma)\, \Big(T_{\perp\perp}\, l^C - T_{\perp s}\,
h^{sr}\, \partial_r\, F^C\Big) (\tau , \sigma^u)\Big]\, \Lambda_C{}^u(\vec
h)\, \Lambda_D{}^v(\vec h).
  \label{2.11}
\end{eqnarray}

\medskip

As shown in Eqs. (5.7) and (5.13) of the first paper in Ref.\cite{2} and in
Eq.(3.4) of Ref.\cite{17}, the three pairs of second class constraints
eliminating the internal center of mass in arbitrary non-inertial rest
frames have the form

\begin{eqnarray}
\mathcal{P}^r &=& \int d^3\sigma\, \sqrt{h(\tau , \sigma^u)}\, \Big[%
T_{\perp\perp}\, l^A - T_{\perp s}\, h^{sr}\, \partial_r\, F^A\Big](\tau ,
\sigma^u)\,\,\, \Lambda_A{}^r(\vec h) \approx 0,  \nonumber \\
&&{}  \nonumber \\
\mathcal{K}^r &=& \int d^{3}\sigma \,\sqrt{h(\tau ,\sigma
^{u})}\,\Big[F^{C}(\tau ,\sigma ^{u})\,\Big(T_{\perp \perp }\,l^{D}-T_{\perp
s}\,h^{sr}\,\partial _{r}\,F^{D}\Big)(\tau ,\sigma ^{u})-\nonumber \\
 &-& F^{D}(\tau ,\sigma
^{u})\,\Big(T_{\perp \perp }\,l^{C}-T_{\perp s}\,h^{sr}\,\partial _{r}\,F^{C}
\Big)(\tau ,\sigma ^{u})\Big]\,\Lambda _{C}{}^{r}(\vec{h})\,\Lambda
_{D}{}^{\tau }(\vec{h}) \approx \nonumber \\
 &\approx& Mc\, h^r\, \Big(x^o_o + f^B(\tau )\,
\Lambda_B{}^C(\vec h)\, \epsilon^o_C(\vec h) - {\frac{{\sum_u\, h^u\, \Big(%
x^u_o - z^u + f^B(\tau )\, \Lambda_B{}^C(\vec h)\, \epsilon^u_C(\vec h)\Big)}%
}{{1 + \sqrt{1 + {\vec h}^2}}}} \Big) -  \nonumber \\
&-& \Big(x^r_o - z^r + f^B(\tau )\, \Lambda_B{}^C(\vec h)\,
\epsilon^r_C(\vec h) + {\frac{{\delta^{rm}\, \epsilon_{mnk}\, h^n\, {\tilde S%
}^k}}{{Mc\, (1 + \sqrt{1 + {\vec h}^2})}}}\Big).
 \label{2.12}
\end{eqnarray}

\medskip

Let us remark that that if we put $\Lambda_A{}^B(\vec h) = \delta^B_A$ and $%
x^{\mu}_o + f^B(\tau)\, \Lambda_B{}^C(\vec h)\, \epsilon^{\mu}_C(\vec h) =
Y^{\mu}(0) + h^{\mu}\, \tau$, then we recover the results for the inertial
rest frame centered on the Fokker-Pryce inertial observer when $F^A(\tau,
\sigma^u) = \sigma^A$.\medskip

Instead the conditions $\Lambda_A{}^B(\vec h) = \delta^B_A$ and $f^B(\tau)\,
\Lambda_B{}^C(\vec h)\, \epsilon^{\mu}_C(\vec h) = h^{\mu}\, \tau$,
identifying the inertial rest frame centered on the inertial observer $%
x^{\mu}_o + h^{\mu}\, \tau$, have the constraints $\mathcal{K}^r \approx 0$
replaced by the second of Eqs.(\ref{2.12}). \medskip

For an inertial frame with $\epsilon^{\mu}_A = \Lambda_A{}^B(\vec h)\,
\epsilon^{\mu}_B(\vec h)$ centered on the inertial observer with world-line $%
x_o^{\mu} + \epsilon_{\tau}^{\mu}\, \tau$ ($\epsilon_{\tau}^{\mu} = l^{\mu}$%
, the normal to the Euclidean 3-space) one has $F^A(\tau, \sigma^u) =
\sigma^A$, $f^A(\tau) = \delta^A_{\tau}\, \tau$, and the second-class
constraints (2.11) but with $\mathcal{P}^r \approx M c h^r = P^r$. \medskip

However, in the non-inertial case it is highly non-trivial to find the
relative variables inside the internal 3-space due to its non-Euclidean
structure. For instance, if we have particles with radar 3-coordinates $
\eta^r_i(\tau)$ in the non-Euclidean 3-space $\Sigma_{\tau}$ and interacting
through action-at-a-distance potentials $V(({\vec \eta}_i(\tau) - {\vec \eta}
_j)^2)$ in the inertial rest frame, their transcription in $\Sigma_{\tau}$
must use a bi-scalar like the Synge world function $\Omega(i,j)$ \cite{18}
built for the space-like 3-geodesic joining the particles $i$ and $j$ in $
\Sigma_{\tau}$ \footnote{
This quantity is a 3-scalar in both points, its gradient with respect the
end points gives the 3-vectors tangent to the 3-geodesic in the end points.
In the Euclidean limit one recovers the quantity $({\vec \eta}_i - {\vec \eta
}_j)^2$.}.

\section{A New Family of Admissible Global Non-Inertial Frames}

Let us now look at a family of admissible global non-inertial frames wider
than the ones quoted in Subsection E of Section II. These new non-inertial
frames will have in general non-Euclidean 3-spaces. They are motivated by
the locality principle \cite{7}, according to which at each instant an
accelerated detector gives the same data as an instantaneously comoving
inertial detector.This inertial detector is connected to a standard
reference inertial frame by a Lorentz transformation whose boost part is
determined by the instantaneous velocity of the accelerated
detector.\medskip

This suggests replacing the embedding (2.1) with the following one ($%
\epsilon _{A}^{\mu }$ asymptotic tetrad, $\epsilon _{A}^{\mu }\,\eta _{\mu
\nu }\,\epsilon _{B}^{\nu }=\eta _{AB}$, $\eta _{\mu \nu }=\epsilon
\,(+---)=\eta _{AB}$; $\sigma =|\vec{\sigma}|=\sqrt{\sum_{r}\,(\sigma
^{r})^{2}}$, $\partial _{r}\,\sigma ={\frac{{\sigma ^{r}}}{{\sigma }}}={\hat{%
\sigma}}^{r}$)

\begin{eqnarray}
z^{\mu}(\tau, \sigma^r) &=& x^{\mu}(\tau) + \epsilon^{\mu}_A\,
\Lambda^A{}_r(\tau, \sigma)\, \sigma^r \, {\rightarrow}_{\sigma \rightarrow
\infty}\, x^{\mu}(\tau) + \epsilon^{\mu}_r\, \sigma^r,  \nonumber \\
&&{}  \nonumber \\
&&{}  \nonumber \\
&&\Lambda^A{}_r(\tau, \sigma)\, {\rightarrow}_{\sigma \rightarrow \infty}\,
\delta^A_r,\qquad \Lambda^A{}_r(\tau, 0)\,\, finite,
 \label{3.1}
\end{eqnarray}

\noindent with the Lorentz matrices $\Lambda^A{}_B(\tau, \sigma)$ satisfying
$\Lambda^A{}_C(\tau, \sigma) \, \eta_{AB}\, \Lambda^B{}_D(\tau, \sigma) =
\eta_{CD}$. \bigskip

The origin of the 3-coordinates $\sigma^r$ in the 3-spaces $\Sigma_{\tau}$
is a time-like (not uniformly accelerated like the Rindler ones) observer
with world-line $x^{\mu}(\tau)$

\begin{eqnarray}
z^{\mu}(\tau, 0) &=& x^{\mu}(\tau) = x^{\mu}_o + \epsilon^{\mu}_A\,
f^A(\tau),  \nonumber \\
&&{}  \nonumber \\
{\dot x}^{\mu}(\tau) &=& \epsilon^{\mu}_A\, {\dot f}^A(\tau) =
\epsilon^{\mu}_A\, \alpha(\tau)\, \gamma_x(\tau)\, \left(
\begin{array}{l}
1 \\
\beta^r_x(\tau)
\end{array}
\right),\qquad \epsilon\, {\dot x}^2(\tau) = \alpha^2(\tau) > 0,  \nonumber
\\
&&{}  \nonumber \\
f^A(\tau) &=& \int_o^\tau\, d\tau_1\, \alpha(\tau_1)\, \gamma_x(\tau_1)\,
\left(
\begin{array}{l}
1 \\
\beta^r_x(\tau)
\end{array}
\right),\qquad \gamma_x(\tau) = {\frac{1}{\sqrt{1 - {\vec \beta}_x^2(\tau)}}}
..  \label{3.2}
\end{eqnarray}

\noindent In this equation ${\vec \beta}_x(\tau)$ ($|{\vec \beta}_x(\tau)| < 1$) is the instantaneous
3-velocity, divided by $c$, of the observer and $\tau$ is the proper time of
the observer when $\alpha(\tau) = 1$.

\bigskip

The Lorentz matrix is parametrized as the product of a boost by a rotation
matrix, $\Lambda^A{}_B(\tau, \sigma)\, =\, B^A{}_C(\tau, \sigma)\, {\tilde R}%
^C{}_B(\tau, \sigma) {\rightarrow}_{\sigma \rightarrow \infty}\, \delta^A_B$%
, with the following notation (R is an ordinary $3 \times 3$ rotation
matrix, $R^{-1} = R^T$; ${\vec {\tilde \beta}}$ is a 3-velocity divided by $%
c $; $\tilde \gamma(\tau, \sigma)\, =\, {\frac{1}{\sqrt{1 - {\vec {\tilde
\beta}}^2(\tau, \sigma)}}}$; $(B^{-1})^A{}_B = B^C{}_D\, \eta_{CB}\,
\eta^{BA}$)

\begin{eqnarray}
B^A{}_C(\tau, \sigma) &=&\left(
\begin{array}{ll}
\tilde \gamma & \tilde \gamma\, {\tilde \beta}^s \\
\tilde \gamma\, {\tilde \beta}^r & \delta^{rs} + {\frac{{{\tilde \gamma}^2\,
{\tilde \beta}^r\, {\tilde \beta}^s}}{{\tilde \gamma + 1}}}%
\end{array}
\right) (\tau, \sigma),  \nonumber \\
&&{}  \nonumber \\
&&{}  \nonumber \\
{\tilde R}^C{}_B(\tau, \sigma) &=& \left(
\begin{array}{ll}
1 & 0 \\
0 & R_{rs}%
\end{array}
\right) (\tau, \sigma),  \nonumber \\
&&{}  \nonumber \\
&&{}  \nonumber \\
\Lambda^A{}_B(\tau, \sigma) &=& \left(
\begin{array}{ll}
\tilde \gamma & \tilde \gamma\, {\tilde \beta}^u\, R_{us} \\
\tilde \gamma\, {\tilde \beta}^r & (\delta^{ru} + {\frac{{{\tilde \gamma}%
^2\, {\tilde \beta}^r\, {\tilde \beta}^u}}{{\tilde \gamma + 1}}} )\, R_{us}%
\end{array}
\right) (\tau, \sigma),
  \label{3.3}
\end{eqnarray}

\noindent The Lorentz matrix become the identity at spatial infinity and is
finite at the origin, where the rotation matrix is assumed to become the
identity and where the 3-velocity may or may not become the one of the
observer, ${\vec {\tilde \beta}}(\tau, 0) = {\vec \beta}_x(\tau)$ or ${\vec {%
\tilde \beta}}(\tau, 0) \not= {\vec \beta}_x(\tau)$.

\bigskip

The angles in the rotation matrix have the same parametrization as in
Subsection E of Section II

\begin{eqnarray}
R_{rs}(\tau ,\sigma ) &=&R_{rs}({\tilde{\alpha}}_{i}(\tau ,\sigma )),\qquad
i=1,2,3,  \nonumber \\
&&{}  \nonumber \\
{\tilde{\alpha}}_{i}(\tau ,\sigma ) &=&f(\sigma )\,\alpha _{i}(\tau ),\qquad
f(\sigma )\,{\rightarrow }_{\sigma \rightarrow \infty }\,0,\quad f(0)=1.
\label{3.4}
\end{eqnarray}

\bigskip

The parameters in the Lorentz boosts have the following parametrization

\begin{eqnarray}
B^A{}_B(\tau, \sigma) &=& B^A{}_B({\tilde \beta}^r(\tau, \sigma)), \qquad r
= 1,2,3,  \nonumber \\
&&{}  \nonumber \\
{\tilde \beta}^r(\tau, \sigma) &=& g(\sigma)\, \beta^r(\tau),\qquad
g(\sigma)\, {\rightarrow}_{\sigma \rightarrow \infty}\, 0, \quad g(0) = 1,
\nonumber \\
&&{}  \nonumber \\
&&\tilde \gamma(\tau, \sigma)\, =\, {\frac{1}{\sqrt{1 - g^2(\sigma)\, {\vec
\beta}^2(\tau)}}}.
 \label{3.5}
\end{eqnarray}

\bigskip

In terms of the quantities (\ref{a1}), (\ref{a2}), defined in Appendix A and
of Eqs. (\ref{a4}), (\ref{a5}), (\ref{a6}), we get the following results for
the metric ${}^4g_{AB}(\tau, \sigma^u) = \Big(z^{\mu}_A\, \eta_{\mu\nu}\,
z^{\nu}_B\Big)(\tau, \sigma^u)$ \medskip

\begin{eqnarray}
\epsilon\, {}^4g_{\tau\tau}(\tau, \sigma^u) &=& \epsilon\, \Big(%
z^{\mu}_{\tau}\, \eta_{\mu\nu}\, z^{\nu}_{\tau}\Big)(\tau, \sigma^u) =
\alpha^2(\tau) +  \nonumber \\
&+&2\, \sigma\, \alpha(\tau)\, \gamma_x(\tau)\, \Big( {\frac{{1 -
g(\sigma)\, {\vec \beta}_x(\tau) \cdot \vec \beta(\tau)}}{\sqrt{1 -
g^2(\sigma)\, {\vec \beta}^2(\tau)}}}\, g(\sigma)\, {\dot {\vec \beta}}%
(\tau) \cdot \sum_n\, {\vec \Omega} _{(B)}{}^{\tau}{}_n\, {\hat \sigma}^n +
\nonumber \\
&+&\sum_{uv}\, \Big[ {\frac{{g(\sigma)\, \beta^u(\tau)}}{{1 + \sqrt{1 -
g^2(\sigma)\, {\vec \beta}^2(\tau)}}}} \, \Big(1 + {\frac{{1 - g(\sigma)\, {%
\vec \beta}_x(\tau) \cdot \vec \beta(\tau)}}{\sqrt{1 - g^2(\sigma)\, {\vec
\beta}^2(\tau)}}}\Big) - \beta^u_x(\tau)\Big]\, R_{uv}  \nonumber \\
&&\Big[f(\sigma)\, {\frac{{\Omega_{(R)}}}{c}}\, (\hat \sigma \times \hat
n)^v + g(\sigma)\, {\dot {\vec \beta}}(\tau) \cdot \sum_m\,{\vec \Omega}%
_{(B)}{}^v{}_m\, {\hat \sigma}^m \Big]\Big)(\tau, \sigma) -  \nonumber \\
&-& \sigma^2\, \Big(f^2(\sigma)\, {\frac{{\Omega^2_{(R)}}}{{c^2}}}\, (\hat
\sigma \times \hat n)^2 +  \nonumber \\
&+& f(\sigma)\, g(\sigma)\, {\frac{{\Omega_{(R)}}}{c}}\, \sum_v\, (\hat
\sigma \times \hat n)^v\, {\dot {\vec \beta}}(\tau) \cdot \sum_n\, {\vec
\Omega}_{(B)}{}^v{}_n\, {\hat \sigma}^n +  \nonumber \\
&+& g^2(\sigma)\, \sum_{nm}\, \Big[{\dot {\vec \beta}}(\tau) \cdot {\vec
\Omega}_{(B)}{}^{\tau}{}_m\, {\dot {\vec \beta}}(\tau) \cdot {\vec \Omega}%
_{(B)}{}^{\tau}{}_n -  \nonumber \\
&-& \sum_v\, {\dot {\vec \beta}}(\tau) \cdot {\vec \Omega}_{(B)}{}^v{}_m\, {%
\dot {\vec \beta}}(\tau) \cdot {\vec \Omega}_{(B)}{}^v{}_n \Big]\, {\hat
\sigma}^m\, {\hat \sigma}^n\Big)(\tau, \sigma).
 \label{3.6}
\end{eqnarray}

\begin{eqnarray}
\epsilon\, {}^4g_{\tau r}(\tau, \sigma^u) &=& \epsilon\, \Big(%
z^{\mu}_{\tau}\, \eta_{\mu\nu}\, z^{\nu}_r\Big)(\tau, \sigma^u) = -
N_r(\tau, \sigma^u) =  \nonumber \\
&=&\alpha(\tau)\, \gamma_x(\tau)\, \Big[\Lambda^{\tau}{}_r - \sum_v\,
\beta^v_x(\tau)\, \Lambda^v{}_r +  \nonumber \\
&+& \sigma\, {\hat \sigma}^r\, \Big(f^{^{\prime }}(\sigma)\, \sum_u\,
[\Lambda^{\tau}{}_u - \sum_v\, \beta^v_x(\tau)\, \Lambda^v{}_u]\, (\hat
\sigma \times {\check \Omega}_{(R)})^u +  \nonumber \\
&+&g^{^{\prime }}(\sigma)\, [\Lambda^{\tau}{}_{\tau} - \sum_v\,
\beta^v_x(\tau)\, \Lambda^v{}_{\tau}]\, \vec \beta(\tau) \cdot \sum_n\, {%
\vec \Omega} _{(B)}{}^{\tau}{}_n\, {\hat \sigma}^n \Big)\Big](\tau, \sigma) -
\nonumber \\
&-& \sigma\, \Big[f(\sigma)\, {\frac{{\Omega_{(R)}}}{c}}\, (\hat \sigma
\times \hat n)^r + g(\sigma)\, {\dot {\vec \beta}}(\tau) \cdot \sum_n\, {%
\vec \Omega}_{(B)}{}^r{}_n\, {\hat \sigma}^n +  \nonumber \\
&+& \sigma\, {\hat \sigma}^r\, \Big(f(\sigma)\, f^{^{\prime }}(\sigma)\, {%
\frac{{\Omega_{(R)}}}{c}}\, (\hat \sigma \times \hat n) \cdot (\hat \sigma
\times {\check \Omega}_{(R)}) +  \nonumber \\
&+& g(\sigma)\, f^{^{\prime }}(\sigma)\, \sum_{vn}\, {\dot {\vec \beta}}%
(\tau) \cdot {\vec \Omega}_{(B)}{}^v{}_n\, {\hat \sigma}^n\, (\hat \sigma
\times {\check \Omega}_{(R)})^v -  \nonumber \\
&-&g(\sigma)\, g^{^{\prime }}(\sigma)\, \sum_{nm}\, {\dot {\vec \beta}}
(\tau) \cdot {\vec \Omega}_{(B)}{}^{\tau}{}_m\, {\vec \beta}(\tau) \cdot {\
\vec \Omega}_{(B)}{}^{\tau}{}_n {\hat \sigma}^m\, {\hat \sigma}^n \Big)\Big]%
(\tau, \sigma),
  \label{3.7}
\end{eqnarray}

\bigskip

\begin{eqnarray}
- \epsilon\, {}^4g_{rs}(\tau, \sigma^u) &=& h_{rs}(\tau, \sigma^u) = -
\epsilon\, \Big(z^{\mu}_r\, \eta_{\mu\nu}\, z^{\nu}_s\Big)(\tau, \sigma^u) =
\nonumber \\
&=& \delta_{rs} + \sigma\, f^{^{\prime }}(\sigma)\, \Big({\hat \sigma}^r\,
(\hat \sigma \times {\check \Omega}_{(R)})^s + {\hat \sigma}^s\, (\hat
\sigma \times {\ \check \Omega}_{(R)})^r \Big)(\tau, \sigma) +  \nonumber \\
&+& \sigma^2\, {\hat \sigma}^r\, {\hat \sigma}^s\, \Big(f^{{^{\prime }}
2}(\sigma)\, (\hat \sigma \times {\check \Omega}_{(R)})^2 -  \nonumber \\
&-& g^{{^{\prime }} 2}(\sigma)\, \sum_{nm}\, {\vec \beta}(\tau) \cdot {\
\vec \Omega}_{(B)}{}^{\tau}{}_m\, {\vec \beta}(\tau) \cdot {\vec \Omega}
_{(B)}{}^{\tau}{}_n \, {\hat \sigma}^m\, {\hat \sigma}^n \Big)(\tau, \sigma).
\nonumber \\
&&{}  \label{3.8}
\end{eqnarray}

\bigskip

From Eq.(\ref{a8}) of Appendix A, after a lengthy calculation, we get the
following expression for the lapse function $N(\tau, \sigma^u) = 1 + n(\tau,
\sigma^u)$

\begin{eqnarray}
\Big(\sqrt{h}\, (1 + n)\Big)(\tau, \sigma^u) &=& \Big[\alpha(\tau)\,
\gamma_x(\tau)\, {\frac{{1 - g(\sigma)\, \vec \beta(\tau) \cdot {\vec \beta}%
_x(\tau)}}{\sqrt{1 - g^2(\sigma)\, {\vec \beta}^2(\tau)}}} +  \nonumber \\
&+& \sigma\, g(\sigma)\, {\dot {\vec \beta}}(\tau) \cdot \sum_n\, {\vec
\Omega}_{(B)}{}^{\tau}{}_n\, {\hat \sigma}^n\Big](\tau, \sigma) -  \nonumber
\\
&-& \sigma\, g^{^{\prime }}(\sigma)\, \vec \beta(\tau) \cdot \sum_n\, {\vec
\Omega}_{(B)}(\tau, \sigma){}^{\tau}{}_n\, {\hat \sigma}^n  \nonumber \\
&&\sum_v\, {\hat \sigma}^v\, \Big[\alpha(\tau)\, \gamma_x(\tau)\, \sum_u\,
R^T_{vu}  \nonumber \\
&&\Big(\beta^u_x(\tau) - {\frac{{g(\sigma)\, \beta^u(\tau)}}{{1 + \sqrt{1 -
g^2(\sigma)\, {\vec \beta}^2(\tau)}}}}\, \Big[1 + {\frac{{1 - g(\sigma)\, {%
\vec \beta}_x(\tau) \cdot \vec \beta(\tau)}}{\sqrt{1 - g^2(\sigma)\, {\vec
\beta}^2(\tau)}}}\Big]\Big) +  \nonumber \\
&+& \sigma\, g(\sigma)\, {\dot {\vec \beta}}(\tau) \cdot \sum_m\, {\vec
\Omega}_{(B)}{}^v{}_m\, {\hat \sigma}^m\Big](\tau, \sigma).
  \label{3.9}
\end{eqnarray}

 \medskip

The positivity requirement for the quantities of Eqs. (3.6) and (3.9)
together with the positivity of the three eigenvalues $\lambda _{i}(\tau
,\sigma ^{u})>0$ of the matrix (3.8) are the conditions on the observer and
on the Lorentz matrix of Eq.(3.1) for having a nice foliation, i.e. a well
defined global non-inertial frame described by the embedding (3.1).
Therefore one must have

\begin{eqnarray}
&&1 + n(\tau, \sigma^u) > 0,\qquad \epsilon\, {}^4g_{\tau\tau}(\tau,
\sigma^u) > 0,  \nonumber \\
{}&&  \nonumber \\
&&\Big(\lambda_1\, \lambda_2\, \lambda_3\Big)(\tau, \sigma^u) = h(\tau,
\sigma^u) > 0,  \nonumber \\
&&\Big(\lambda_1 + \lambda_2 + \lambda_3\Big)(\tau, \sigma^u) = \Big(h_{11}
+ h_{22} + h_{33}\Big)(\tau, \sigma^u) > 0,  \nonumber \\
&&\Big(\lambda_1\, \lambda_2 + \lambda_2\, \lambda_3 + \lambda_3\, \lambda_1%
\Big)(\tau, \sigma^u) = \Big( h_{11}\, h_{22} - h_{12}\, h_{21} +  \nonumber
\\
&&+ h_{22}\, h_{33} - h_{23}\, h_{32} + h_{33}\, h_{11} - h_{13}\, h_{31} %
\Big)(\tau, \sigma^u) > 0.
  \label{3.10}
\end{eqnarray}

\subsection{Some Solutions to the Positivity Requirements.}

By using Eqs.(A2) we get the following explicit expressions for Eqs. (3.6),
(3.8) and (3.9)

\begin{eqnarray*}
\epsilon\, {}^4g_{\tau\tau}(\tau, \sigma^u) &=& \alpha^2(\tau) + 2 \sigma\,
\alpha(\tau)\, \gamma_x(\tau)\, \Big( {\frac{{1 - g(\sigma)\, {\vec \beta}%
_x(\tau) \cdot \vec \beta(\tau)}}{\sqrt{1 - g^2(\sigma)\, {\vec \beta}%
^2(\tau)}}}\, g(\sigma)\, \sum_{su}\, {\dot \beta}^s(\tau)\, R_{su}(\tau,
\sigma)\, {\hat \sigma}^u +  \nonumber \\
&+& \sum_{uv}\, \Big[ {\frac{{g(\sigma)\, \beta^u(\tau)}}{{1 + \sqrt{1 -
g^2(\sigma)\, {\vec \beta}^2(\tau)}}}} \, \Big(1 + {\frac{{1 - g(\sigma)\, {%
\vec \beta}_x(\tau) \cdot \vec \beta(\tau)}}{\sqrt{1 - g^2(\sigma)\, {\vec
\beta}^2(\tau)}}}\Big) - \beta^u_x(\tau)\Big]\, R_{uv}(\tau, \sigma)
\nonumber \\
&&\Big[f(\sigma)\, {\frac{{\Omega_{(R)}(\tau, \sigma)}}{c}}\, (\hat \sigma
\times \hat n(\tau, \sigma))^v +  \nonumber \\
&+&g^2(\sigma)\, \sum_{mn}\, R^T_{vm}(\tau, \sigma)\, ({\dot \beta}%
^m(\tau)\, \beta^n(\tau) - {\dot \beta}^n\, \beta^m(\tau))\, R_{sn}(\tau,
\sigma)\, {\hat \sigma}^n \Big] \Big) -  \nonumber \\
&-& \sigma^2\, \Big(f^2(\sigma)\, {\frac{{\Omega^2_{(R)}(\tau, \sigma)}}{c^2}%
}\, (\hat \sigma \times \hat n(\tau, \sigma))^2 +  \nonumber \\
&+& f(\sigma)\, g^2(\sigma)\, {\frac{{\Omega_{(R)}(\tau, \sigma)}}{c}}\,
\sum_{vmnu}\, (\hat \sigma \times \hat n(\tau, \sigma))^v  \nonumber \\
&&R^T_{vm}(\tau, \sigma)\, ({\dot \beta}^m(\tau)\, \beta^n(\tau) - {\dot
\beta}^n\, \beta^m(\tau))\, R_{nu}(\tau, \sigma)\, {\hat \sigma}^u +
\nonumber \\
&+& g^2(\sigma)\, \sum_{mn}\, \Big[\sum_{rs}\, {\dot \beta}^s(\tau)\,
R_{rm}(\tau, \sigma)\, {\dot \beta}^s(\tau)\, R_{sn}(\tau, \sigma) -
\nonumber \\
&-& \sum_{vursw}\, R^T_{vu}(\tau, \sigma)\, ({\dot \beta}^u(\tau)\,
\beta^r(\tau) - {\dot \beta}^r(\tau)\, \beta^u(\tau))\, R_{rm}(\tau, \sigma)
\nonumber \\
&&R^T_{vw}(\tau, \sigma)\, ({\dot \beta}^w(\tau)\, \beta^s(\tau) - {\dot
\beta}^s(\tau)\, \beta^w(\tau))\, R_{sn}(\tau, \sigma)\Big]\, {\hat \sigma}%
^m\, {\hat \sigma}^n\Big) > 0,
\end{eqnarray*}

\begin{eqnarray*}
h_{rs}(\tau, \sigma^u) &=& \delta_{rs} + \sigma\, f^{^{\prime }}(\sigma)\, %
\Big({\hat \sigma}^r\, (\hat \sigma \times {\check \Omega}_{(R)}(\tau,
\sigma))^s + {\hat \sigma}^s\, (\hat \sigma \times {\ \check \Omega}%
_{(R)}(\tau, \sigma))^r \Big) +  \nonumber \\
&+& \sigma^2\, {\hat \sigma}^r\, {\hat \sigma}^s\, \Big[f^{{^{\prime }}
2}(\sigma)\, (\hat \sigma \times {\check \Omega}_{(R)}(\tau, \sigma))^2 -
\nonumber \\
&-& g^{{^{\prime }} 2}(\sigma)\, \Big(\sum_{nm}\, \beta^n(\tau)\,
R_{nm}(\tau, \sigma)\, {\hat \sigma}^m \Big)^2\Big],
\end{eqnarray*}

\begin{eqnarray}
\sqrt{h(\tau, \sigma^u)}\,(1 + n(\tau, \sigma^u)) &=& \alpha(\tau)\,
\gamma_x(\tau)\, {\frac{{1 - g(\sigma)\, \vec \beta(\tau) \cdot {\vec \beta}%
_x(\tau)}}{\sqrt{1 - g^2(\sigma)\, {\vec \beta}^2(\tau)}}} + \sigma\,
g(\sigma)\, \sum_{sn}\, {\dot \beta}^s(\tau)\, R_{sn}(\tau, \sigma)\, {\hat
\sigma}^n -  \nonumber \\
&-& \sigma\, g^{^{\prime }}(\sigma)\, \sum_{snv}\, \beta^s(\tau)\,
R_{sn}(\tau, \sigma)\, {\hat \sigma}^n\, {\hat \sigma}^v\, \Big[%
\alpha(\tau)\, \gamma_x(\tau)\, \sum_u\, R^T_{vu}(\tau, \sigma)  \nonumber \\
&&\Big(\beta^u_x(\tau) - {\frac{{g(\sigma)\, \beta^u(\tau)}}{{1 + \sqrt{1 -
g^2(\sigma)\, {\vec \beta}^2(\tau)}}}}\, \Big[1 + {\frac{{1 - g(\sigma)\, {%
\vec \beta}_x(\tau) \cdot \vec \beta(\tau)}}{\sqrt{1 - g^2(\sigma)\, {\vec
\beta}^2(\tau)}}}\Big]\Big) +  \nonumber \\
&+& \sigma\, g^2(\sigma)\, \sum_{uwr}\, R^T_{vu}(\tau, \sigma)\, ({\dot \beta%
}^u(\tau)\, \beta^w(\tau) - {\dot \beta}^w(\tau)\, \beta^u(\tau))\,
R_{wr}(\tau, \sigma)\, {\hat \sigma}^r\Big] > 0.  \nonumber \\
&&{}  \label{3.11}
\end{eqnarray}

The positivity conditions are the restrictions on the form factors $%
f(\sigma) $ and $g(\sigma)$ implying that the global non-inertial frame with
its non-Euclidean 3-spaces is well defined, i.e. it has no pathology.

\medskip

Due to the complicated form of the positivity conditions, we give explicitly
only two special families of solutions.

\subsubsection{Boosts with Small Velocities}

When the boost parameter $\beta ^{r}(\tau )$ of Eq.(3.5) and its time
variation ${\dot{\beta}}^{r}(\tau )$ are small quantities of order $\epsilon
$ ($|\vec{\beta}(\tau )|,|{\dot{\vec{\beta}}}(\tau )| \approx O(\epsilon )<<1$;
nearly non-relativistic small velocities), Eqs.(\ref{3.11}) become

\begin{eqnarray*}
\epsilon\, {}^4g_{\tau\tau}(\tau, \sigma^u) &=& \alpha^2(\tau) + 2 \sigma\,
\alpha(\tau)\, \gamma_x(\tau)\, f(\sigma)\, {\frac{{\Omega_{(R)}(\tau,
\sigma)}}{c}}\, \sum_{uv}\, \beta^u_x(\tau)\, R_{uv}(\tau, \sigma)\, (\hat
\sigma \times \hat n(\tau, \sigma))^v -  \nonumber \\
&-& \sigma^2\, f^2(\sigma)\, {\frac{{\Omega^2_{(R)}(\tau, \sigma)}}{c^2}}\,
(\hat \sigma \times \hat n(\tau, \sigma))^2 + O(\epsilon) > 0,  \nonumber \\
{}&&  \nonumber \\
\sqrt{h(\tau, \sigma^u)}\,(1 + n(\tau, \sigma^u)) &=& \alpha(\tau)\,
\gamma_x(\tau) + O(\epsilon) > 0,
\end{eqnarray*}

\begin{eqnarray}
h_{rs}(\tau, \sigma^u) &=& \delta_{rs} + \sigma\, f^{^{\prime }}(\sigma)\, %
\Big({\hat \sigma}^r\, (\hat \sigma \times {\check \Omega}_{(R)}(\tau,
\sigma))^s + {\hat \sigma}^s\, (\hat \sigma \times {\ \check \Omega}%
_{(R)}(\tau, \sigma))^r \Big) +  \nonumber \\
&+& \sigma^2\, {\hat \sigma}^r\, {\hat \sigma}^s\, f^{{^{\prime }}
2}(\sigma)\, (\hat \sigma \times {\check \Omega}_{(R)}(\tau, \sigma))^2 +
O(\epsilon^2).
  \label{3.12}
\end{eqnarray}

Eqs.(3.12) are the conditions for differentially rotating non-inertial
frames in Euclidean 3-spaces. Now the 3-spaces have deviations of order $%
O(\epsilon)$ from Euclidean 3-spaces and there is no restriction on $%
g(\sigma)$. Instead $f(\sigma)$ must satisfy Eq.(2.9) and this also implies
the positivity of the three eigenvalues of the 3-metric.

\subsubsection{Time-Independent Boosts}

Let us now consider time-independent boosts: ${\dot {\vec \beta}}(\tau) = 0$%
, i.e. $\vec \beta (\tau) = \vec \beta = const.$ At every time there is the
same non-Euclidean 3-space. Since Eqs.(3.11) remain complicated, let us also
put the restriction $\vec \beta \cdot {\vec \beta}_x(\tau) = 0$ on the
world-line $x^{\mu}(\tau)$ of the observer. Then Eqs.(3.11) become

\begin{eqnarray*}
\epsilon\, {}^4g_{\tau\tau}(\tau, \sigma^u) &=& \alpha^2(\tau) + 2 \sigma\,
\alpha(\tau)\, \gamma_x(\tau)\, f(\sigma)\, {\frac{{\Omega_{(R)}(\tau,
\sigma)}}{c}}  \nonumber \\
&& \sum_{uv}\, \Big(\beta^u_x(\tau) - {\frac{{g(\sigma)\, \beta^u}}{\sqrt{1
- g^2(\sigma)\, {\vec \beta}^2}}}\Big)\, R_{uv}(\tau, \sigma)\, (\hat \sigma
\times \hat n(\tau, \sigma))^v -  \nonumber \\
&-& \sigma^2\, f^2(\sigma)\, {\frac{{\Omega^2_{(R)}(\tau, \sigma)}}{c^2}}\,
(\hat \sigma \times \hat n(\tau, \sigma))^2 > 0,
\end{eqnarray*}

\begin{eqnarray*}
h_{rs}(\tau, \sigma^u) &=& \delta_{rs} + \sigma\, f^{^{\prime }}(\sigma)\, %
\Big({\hat \sigma}^r\, (\hat \sigma \times {\check \Omega}_{(R)}(\tau,
\sigma))^s + {\hat \sigma}^s\, (\hat \sigma \times {\ \check \Omega}%
_{(R)}(\tau, \sigma))^r \Big) +  \nonumber \\
&+& \sigma^2\, {\hat \sigma}^r\, {\hat \sigma}^s\, \Big[f^{{^{\prime }}
2}(\sigma)\, (\hat \sigma \times {\check \Omega}_{(R)}(\tau, \sigma))^2 -
\nonumber \\
&-& g^{{^{\prime }} 2}(\sigma)\, \Big(\sum_{nm}\, \beta^n\, R_{nm}(\tau,
\sigma)\, {\hat \sigma}^m \Big)^2\Big],
\end{eqnarray*}

\begin{eqnarray}
\sqrt{h(\tau, \sigma^u)}\,(1 + n(\tau, \sigma^u)) &=& \alpha(\tau)\,
\gamma_x(\tau)\, \Big[ {\frac{1}{\sqrt{1 - g^2(\sigma)\, {\vec \beta}^2}}} -
\nonumber \\
&-& \sigma\, g^{^{\prime }}(\sigma)\, \sum_{snvu}\, \beta^s\, R_{sn}(\tau,
\sigma)\, {\hat \sigma}^n\, {\hat \sigma}^v\, R^T_{vu}(\tau, \sigma)\, \Big(%
\beta^u_x(\tau) - {\frac{{g(\sigma)\, \beta^u}}{\sqrt{1 - g^2(\sigma)\, {%
\vec \beta}^2}}}\Big) \Big] > 0.  \nonumber \\
&&{}  \label{3.13}
\end{eqnarray}

If $|{\vec \beta}^2|$ is small, the 3-metric $h_{rs}(\tau, \sigma^u)$ has
small deviations from the pure rotational case and its three eigenvalues
remain positive if $f(\sigma)$ satisfies Eq.(2.9). Moreover if for every $%
\tau$ and $\sigma$ we have the following restriction on $g(\sigma)$

\begin{equation}
\beta^u_x(\tau) > {\frac{{g(\sigma)\, \beta^u}}{\sqrt{1 - g^2(\sigma)\, {%
\vec \beta}^2}}},  \label{3.14}
\end{equation}

\noindent then also the condition $\epsilon\, {}^4g_{\tau\tau}(\tau, \sigma)
> 0$ is satisfied by $f(\sigma)$ of Eq.(2.9) with a different constant $A^2$%
.. \medskip

Finally the condition $1 + n(\tau, \sigma^u) > 0$ implies

\begin{eqnarray}
&&- {\frac{{C^2}}{{\sigma}}} < g^{^{\prime }}(\sigma) < {\frac{{C^2}}{{\sigma%
}}},  \nonumber \\
{}&&  \nonumber \\
&&C^2 = min_{\tau, \sigma}\, |\sum_{snvu}\, \beta^s\, R_{sn}(\tau, \sigma)\,
{\hat \sigma}^n\, {\hat \sigma}^v\, R^T_{vu}(\tau, \sigma)\, \Big(\sqrt{1 -
g^2(\sigma)\, {\vec \beta}^2}\, \beta^u_x(\tau) - g(\sigma)\, \beta^u\Big)|.
\label{3.15}
\end{eqnarray}

In conclusion we now have control on some families of global non-inertial
frames with non-Euclidean 3-spaces. When there will be experimental reasons
for studying other families of such frames, one will deepen the study of
Eqs.(3.11).

\section{Comparison of Inertial and non-inertial reference frames centered on
either mathematical or dynamical observers: dynamical inertial Alice versus
dynamical non-inertial Bob}

Both in Galilei and Minkowski space-times the description of isolated
systems is ideally done by an either inertial or accelerated observer origin
of an either inertial or non-inertial frame. The primary role of these
"mathematical" observers is to define a 4-coordinate system and a system of
axes (a tetrad). They are considered as mathematical idealizations of
"dynamical" observers (Alice, Bob, Charlie,..) endowed with macroscopic
apparatuses with which they perform measurements on the given isolated
system (breaking its isolation) with a subsequent mutual communication of
the results obtained. But already at the classical level this transition
from mathematical to dynamical observers requires the inclusion of the
observers in the isolated system to have control of the either Galilei or
Poincar\'{e} generators (so that, for instance, the total energy of the new
system is finite) and to have the world-lines of the observers dynamically
determined (and not given by hand like with mathematical observers). In SR
the spatial non-separability due to the elimination of relative times and
the non-measurability of the relativistic center of mass induce an ever
bigger difference between mathematical and dynamical observers. \medskip

This type of problems becomes extremely complicated at the quantum level due
to its unsolved foundational problems (interpretation; theory of
measurement) and due to the absence of a notion of "reality" of quantum
systems (see Ref. \cite{8} for a discussion of these problems in the
framework presented in this paper). Usually the quantum system is described
with respect to an inertial reference frame centered on a classical
mathematical observer; this observer either carries or describes the
location and orientation of a measuring apparatus. A dynamical observer
should be identified with such an apparatus considered as either a
macroscopic classical object or quantum system (with some semi-classical
description of its quantum many-body structure). As a consequence a
distinction between macroscopic dynamical observers and a microscopic
quantum system becomes extremely problematic already at the non-relativistic
level even before taking into account the spatial non-separability of SR
(see for instance Ref.\cite{19}).

\bigskip

Let us come back to classical SR. Having found families of admissible 3+1
splittings (global non-inertial frames) of Minkowski space-time centered on
arbitrary time-like mathematical observers, we can face the following two
problems:\medskip

A) how to compare the description of an isolated system given by two
different mathematical observers and in particular how to rewrite all the
results which can be obtained in an inertial rest frame in an arbitrary
non-inertial frame;\medskip

B) whether it is possible to get the description of an isolated system in a
non-inertial rest frame centered on a particle of the system used as a
\textit{dynamical} observer.

\subsection{Alice and Bob: Accelerated Mathematical Observers}

Let us consider the two world-lines $x_1^{\mu}(\tau)$ and $x_2^{\mu}(\tau)$
of two time-like observers (Alice and Bob) in Minkowski space-time in a
given inertial frame with Cartesian coordinates $x^{\mu}$ centered on a
mathematical inertial observer. \medskip

To each observer we can associate a non-inertial frame giving two nice
foliations centered on the two observers and defining the radar
4-coordinates $(\tau _{1},\sigma _{1}^{r})$ and $(\tau _{2},\sigma _{2}^{r})$
for parametrizing their respective 3-spaces. There will be two embeddings $%
x^{\mu }=z_{1}^{\mu }(\tau _{1},\sigma _{1}^{r})$ and $x^{\mu }=z_{2}^{\mu
}(\tau _{2},\sigma _{2}^{r})$ for the 3-spaces of the two foliations with
the following equations allowing one to express one set of radar
4-coordinates in terms of the other

\begin{eqnarray}
x^{\mu} &=& z_1^{\mu}(\tau_1, \sigma_1^r) = z_2^{\mu}(\tau_2, \sigma_2),
\nonumber \\
&&\Downarrow  \nonumber \\
&&  \nonumber \\
&&\tau_1 = f_1(\tau_2, \sigma^u_2),\qquad \sigma_1^r = f_1^r(\tau_2,
\sigma^u_2),  \nonumber \\
&&\tau_2 = f_2(\tau_1, \sigma^u_1),\qquad \sigma_2^r = f_2^r(\tau_1,
\sigma^u_1).  \label{4.1}
\end{eqnarray}

\noindent Therefore we get the following identification of the two
world-lines \footnote{$\eta^u_2(\tau_1)$ are the 3-coordinates of $x^{\mu}_2(\tau)$ if we use the foliation of $x^{\mu}_1(\tau)$; ${\tilde \eta}^u_1(\tau_2)$ are the 3-coordinates of $x^{\mu}_1(\tau)$ if we use the foliation of $x^{\mu}_2(\tau)$.}

\begin{eqnarray}
&& x_1^{\mu}(\tau_1) = z_1^{\mu}(\tau_1, 0) = z_2^{\mu}(\tau_2, {\tilde \eta}%
^u_1(\tau_2)) = {\tilde x}_1^{\mu}(\tau_2) = x_1^{\mu}(\tau_1 = f_1(\tau_2, {%
\tilde \eta}^u_1(\tau_2))),  \nonumber \\
&&  \nonumber \\
&& x_2^{\mu}(\tau_1) = z_1^{\mu}(\tau_1, \eta^u_2(\tau_1)) =
z_2^{\mu}(\tau_2, 0) = {\tilde x}_2^{\mu}(\tau_2) = {\tilde x}%
_2^{\mu}(\tau_2 = f_2(\tau_1, \eta_2^u(\tau_1))).  \label{4.2}
\end{eqnarray}

\bigskip

If we have an isolated system of N dynamical particles, their world-lines
will have the following expression

\begin{eqnarray}
y_i^{\mu}(\tau_1) &=& z_1^{\mu}(\tau_1, \eta^u_{yi}(\tau_1)) =
z_2^{\mu}(\tau_2, {\tilde \eta}^u_{yi}(\tau_2)) ={\tilde y}^{\mu}_i(\tau_2)
= z_2^{\mu}(f_2(\tau_1, \eta^u_{yi}(\tau_1)), {\tilde \eta}^u_2(f_2(\tau_1,
\eta^u_{yi}(\tau_1)))),  \nonumber \\
&&{}  \nonumber \\
{\tilde \eta}^r_{yi}(\tau_2) &=& {\tilde \eta}^r_{yi}(f_2(\tau_1,
\eta^u_{yi}(\tau_1))) = \eta^r_{yi}(\tau_1) = \eta^r_{yi}(f_1(\tau_2, {%
\tilde \eta}^u_{yi}(\tau_2))),\qquad i=1,..,N,  \label{4.3}
\end{eqnarray}

\noindent in the two non-inertial frames. \medskip

In general the 3-spaces $\Sigma_{\tau_1}$ and $\Sigma_{\tau_2}$ of the two
foliations will intersect each other, since they correspond to different
clock synchronization conventions. Therefore, in general the two foliations
associated with the two observers do not have a common 3-space to be used as
a common Cauchy surface. We can only transcribe the solutions of the
equations of motion of an isolated system with Cauchy data on a 3-space of
observer 1 in the radar 4-coordinates of observer 2 (or viceversa).

\medskip

As said in Subsection F of the Introduction the second-class constraints
eliminating the internal center of mass and the form of the relative
variables are very complicated, so that it is very difficult to rewrite
Eqs.(4.3) in terms of the Jacobi data $\vec z$, $\vec h$ (the same for Alice
and Bob) of the external center of mass and of the relative variables.

\subsection{Alice Inertial and Bob Accelerated Mathematical Observers}

Let observer 1 (Alice) be in the inertial rest frame with embedding $%
z_{1}^{\mu }(\tau _{1},\sigma _{1}^{r})=z_{W}^{\mu }(\tau _{1},\sigma
_{1}^{r})=Y^{\mu }(\tau _{1})+\epsilon _{r}^{\mu }(\vec{h})\,\sigma _{1}^{r}$%
.. Instead the observer 2 (Bob) is the origin of the non-inertial frame with
given embedding $z_{2}^{\mu }(\tau _{2},\sigma _{2}^{u})$ and has the
world-line $x_{2}^{\mu }(\tau _{2})=z_{2}^{\mu }(\tau _{2},0)$. The
world-line of Alice is the Fokker-Planck center of inertia, $x_{1}^{\mu
}(\tau _{1})=Y^{\mu }(\tau _{1})$:  in its expression (2.3) $M$ and $\vec{S}$ are the
mass and the spin of the isolated system expressed in terms of the relative
variables of Alice by means of Eqs.(2.6).\medskip

Since Eq.(2.3) implies $Y^{\mu}(\tau_1) + \epsilon^{\mu}_r(\vec h)\,
\sigma^r_1 = h^{\mu}\, \tau_1 + \epsilon_r^{\mu}(\vec h)\, \Big(\sigma_1^r +
{\frac{{z^r}}{{M c}}} + {\frac{{h^r\, \vec h \cdot \vec z + (\vec S \times
\vec h)^r}}{{M c\, (1 + \sqrt{1 + {\vec h}^2})}}}\Big)$, Eq.(4.1) can be
written in the form $z_W^{\mu}(\tau_1, \sigma_1^r) = z_2^{\mu}(\tau_2,
\sigma_2^u)\, {\buildrel {def}\over {=}}\, h^{\mu}\, {\bar z}_2(\tau_2,
\sigma_2^u) + \epsilon_r^{\mu}(\vec h)\, {\bar z}_2^r(\tau_2, \sigma_2^u)$.
Then we get

\begin{eqnarray}
\tau_1 &=& {\bar z}_2(\tau_2, \sigma_2^u) = \epsilon\, h_{\mu}\,
z_2^{\mu}(\tau_2, \sigma^u_2),  \nonumber \\
\sigma_1^r &=& {\bar z}_2^r(\tau_2, \sigma_2^u) - {\frac{{z^r}}{{M c}}} - {%
\frac{{h^r\, \vec h \cdot \vec z + (\vec S \times \vec h)^r}}{{M c\, (1 +
\sqrt{1 + {\vec h}^2})}}} =  \nonumber \\
&=& \epsilon^r_{\mu}(\vec h)\, z_2^{\mu}(\tau_2, \sigma_2^u) - {\frac{{z^r}}{%
{M c}}} - {\frac{{h^r\, \vec h \cdot \vec z + (\vec S \times \vec h)^r}}{{M
c\, (1 + \sqrt{1 + {\vec h}^2})}}}.  \label{4.4}
\end{eqnarray}

As a consequence for the world-lines $y^{\mu}_i(\tau_1) = Y^{\mu}(\tau_1) +
\epsilon^{\mu}_r(\vec h)\, \eta^r_i(\tau_1)$ of the N particles of the
isolated system we get

\begin{eqnarray}
\Rightarrow&& \eta_i^r(\tau_1) = {\bar z}^r_2(\tau_2, {\tilde \eta}%
_i^u(\tau_2)) - {\frac{{h^r\, \vec h \cdot \vec z + (\vec S \times \vec h)^r}%
}{{M c\, (1 + \sqrt{1 + {\vec h}^2})}}}.  \nonumber \\
&&{}  \label{4.5}
\end{eqnarray}

\medskip

Therefore all the results in the inertial rest frame centered on Alice with
radar 4-coordinates $(\tau_1, \sigma^r_1)$ can be rewritten in the
accelerated frame centered on the accelerated observer 2 (Bob) using the
radar 4-coordinates $(\tau_2, \sigma^r_2)$ of the non-inertial frame.

\bigskip

Let us remark that, as shown in Ref.\cite{2}, the equations of motion for
the matter of the isolated system are very complicated in non-inertial
frames due to the presence of the relativistic inertial forces. The results
of this Subsection allow to avoid the study of the equations of motion in
non-inertial frames: the solution of these equations can be recovered from
the solution of the equations of motion in the inertial rest frame, where
the 3-space is Euclidean and there are not inertial forces (in general
relativity this is not possible). The only problem is to solve the inertial
equations of motion with inertial rest-frame Cauchy data and to impose the
second-class constraints eliminating the internal center of mass.

\subsection{Alice and Bob Dynamical Observers}

We can now take as the world-line of the accelerated observer 2 (Bob) the
solution $x_2^{\mu}(\tau_1)$ of the equations of motion of a dynamical
particle of an isolated N-particle system described by the mathematical
observer 1 (i.e. $x_2^{\mu}(\tau_1) = y_i^{\mu}(\tau_1)$ for some $i$): in
this way we can get the description of the physics from the point of view of
a dynamical observer which is always accelerated. \medskip

If $x_3^{\mu}(\tau_1) = y_j^{\mu}(\tau_1)$ with $j \not= i$ is the
world-line of another dynamical particle, we can also get the description
from the point of view of this second dynamical observer (Charlie).

\bigskip

Therefore, by using the inertial observer Alice to solve the equations of
motion, we can get the description of the same physics given by two
dynamical observers (Bob and Charlie) and we can compare their descriptions
by using Eqs.(\ref{4.1}).

\subsection{Unruh-DeWitt Detectors}

There is a big literature concerning the uniformly accelerated Rindler
observers in connection with the Unruh radiation and the entanglement of
field modes (see the review in Ref. \cite{1}). These observers are not
considered in our framework because their world-lines are asymptotically
tangent to a light-cone at $\tau = \pm \infty$ and therefore they disappear
with the light-cone in the non-relativistic limit (only time-like observers
become Newtonian observers). \medskip

Many papers in this area \cite{20} study global field-mode entanglement by
using either point-like inertial or non-inertial Unruh-DeWitt detectors. The
simplest Unruh-DeWitt detector is a two-level atom with a hybrid
description: a) it moves along a \textit{classical} given world-line and b)
it has quantum interactions with a field implying transitions between the
two levels.

\medskip

In Ref. \cite{21} we gave a pseudo-classical description of a relativistic
two-level atom in the inertial rest frame and its quantization. Therefore either Alice or Bob (or both)
can be described as a relativistic two-level atom interacting with the other
matter components of the isolated system and we can have (and compare) the
description given by two two-level atom dynamical observers.

\section{Comments on Relativistic Quantum Mechanics in Non-Inertial Frames}

After a review of relativistic quantum mechanics (RQM) in the inertial rest
frame we make some comments on how to extend it to non-inertial frames and
on the problem of what could be the meaning of a quantum observer.

\subsection{Relativistic Quantum Mechanics in the Inertial Rest Frame}

In Ref.\cite{4} there is a consistent formulation of RQM of an isolated
system of scalar particles in the inertial rest frame with the correct
non-relativistic limit. As it is shown in this paper one must quantize the
canonically conjugate frozen Jacobi data $\vec z$, $\vec h$, of the external
center of mass and the set of relative variables and relative momenta in the
Wigner-covariant Euclidean 3-space after the elimination of the internal
center of mass with the second-class constraints corresponding to the
rest-frame conditions (see Eq.(\ref{2.7}) for a two-body case). The solution
of the problem of the relativistic collective variables (leading to the
non-local and non-measurable notion of relativistic center of mass), the
elimination of the relative times in relativistic bound states and the
avoidance of causality problems (like the instantaneous spreading of
wave-packets shown by the Hegerfeldt theorem) imply a \textit{spatial
non-separability} according to which the only allowed presentation of the
Hilbert space is $H = H_{HJcom} \otimes H_{rel}$. While $H_{rel}$ is the
Hilbert space of relative variables, $H_{HJcom}$ is the Hilbert space of the
frozen external center of mass \footnote{%
In the non-relativistic limit three presentations are unitarily equivalent: $%
H = H_1 \otimes H_2 \otimes ... = H_{com} \otimes H_{rel} = H_{HJcom}
\otimes H_{rel}$. While $H_i$ are the Hilbert spaces of the single particles
(non-relativistic separability of subsystems as the zeroth postulate of
quantum mechanics) and $H_{com}$ is the Hilbert space of the Newtonian
center of mass, $H_{HJcom}$ is the Hilbert space of its frozen Jacobi data
obtained with a Hamilton-Jacobi transformation. At the relativistic level
the presentation $H_1 \otimes H_2 \otimes ...$ is forbidden by the problem
of relative times, while the presentation $H_{com} \otimes H_{rel}$ has the
causality problems of the Hegerfeldt theorem.}. The Hamiltonian is the
quantum version of the invariant rest mass $M$. \medskip

Due to the non-local non-measurable nature of the relativistic center of
mass, there is the open problem of the type of operator to be used in $%
H_{HJcom} $ for the Jacobi position $\vec z = Mc\, {\vec x}_{NW}(0)$. See
Ref.\cite{8} for a discussion of the localization problems in RQM and on the
possibility that the center-of-mass position operator be a non-self-adjoint
operator. In that paper there is also a discussion on relativistic
entanglement and on the implications of the spatial non-separability
forbidding the identification of subsystems.

\medskip

Since the non-separability is due to the elimination of the internal center
of mass in the 3-space by means of the three pairs of second-class
constraints implementing the rest-frame conditions, one could start with a
quantization with a separable un-physical Hilbert space $H_{unphy} =
H_{HJcom} \otimes {\bar H}_1 \otimes {\bar H}_2 \otimes ...$, where ${\bar H}%
_i$ is the Hilbert space of particle $i$ described by the Wigner-covariant
3-vectors ${\vec \eta}_i(\tau)$, ${\vec \kappa}_i(\tau)$. Then the reduction
to the physical Hilbert space $H = H_{HJcom} \otimes H_{rel}$ could be
realized by imposing a quantum version of the second-class constraints by
means of the Gupta-Bleuler method, namely by selecting as physical states
those vectors in $H_{unphy}$ which imply a vanishing expectation value for
the three pairs of quantum operators corresponding to the second-class
constraints. Even if for free particles the procedure works, there is the
possibility that in general it leads to an inequivalent quantization. Also
the determination of the physical scalar product is not trivial in this
case. \medskip

This procedure seems the most useful one to extend the rest-frame RQM to
arbitrary inertial frames by restricting to inertial frames the second-class
constraints discussed in Subsection F of Section II (see Eqs.(2.11)) for the
elimination of the inner center of mass of general non-inertial frames.

\subsection{Relativistic Quantum Mechanics in Non-Inertial Frames}

To extend RQM to non-inertial frames is a highly non trivial problem. Here
we list some possibilities: \medskip

A) Even if the 3-spaces are in general non-Euclidean, let us assume that we
have found a canonical basis of relative variables in the given non-inertial
frame. Then, modulo ordering problems in the terms containing the
interactions and using as Hamiltonian the inertial rest mass $M$ augmented
by suitable inertial potentials (see Eq.(5.32) of the first paper in Ref.%
\cite{2} for the case of non-inertial rest frames), one can make a
quantization like in the case of the inertial rest frame. However, besides
the problem of finding the scalar product, there is the generic problem that
the physical Hilbert spaces corresponding to different non-inertial frames
could be non unitarily equivalent, namely one could have inequivalent
quantizations for different non-inertial frames, maybe also inequivalent to
the rest-frame RQM.

\medskip

B) Like in the previous Subsection, one could make an unphysical
quantization of the particle 3-variables and then impose the quantum
second-class constraints eliminating the internal center of mass with the
Gupta-Bleuler method. Again there are the problems of the physical scalar
product and of the possibility of inequivalent quantizations.

\medskip

C) Since at the classical level the descriptions in different non-inertial
frames are gauge equivalent in the framework of parametrized Minkowski
theories as said in Subsection B of Section II, one could think of making a quantization
preserving the gauge equivalence (replaced with some type of unitary
equivalence) at the quantum level. This was done in the first paper of Ref.%
\cite{10} (with the non-relativistic limit studied in the second paper) for
the case of Euclidean 3-spaces with the rotating coordinates of Eq.(\ref{2.10}
) by means of the multi-temporal quantization scheme of Refs. \cite{22}.
\medskip

While in A) and B) one quantizes only matter after having fixed the
embedding $z^{\mu}(\tau, \sigma^r)$ (a gauge variable) to a given function
identifying a well defined non-inertial frame (first reduce, then quantize),
now we consider the enlarged phase space containing the matter and the
conjugate variables $z^{\mu}(\tau, \sigma^r)$, $\rho_{\mu}(\tau, \sigma^r)$.
In this phase space there are the four first-class constraints (\ref{2.3}),
i.e. $\mathcal{H}_{\mu}(\tau, \sigma^r) \, {\buildrel {def}\over {=}}\,
\rho_{\mu}\tau, \sigma^r) - \mathcal{G}_{\mu}(\tau, \sigma^r) \approx 0$
(with $\mathcal{G}_{\mu}$ depending upon both the matter and the embedding),
implying that the embeddings $z^{\mu}(\tau, \sigma^r)$ are gauge variables,
besides the second-class ones eliminating the internal center of mass.

\medskip

The idea of the multi-temporal quantization is to quantize only the physical
degrees of freedom of the particles, but not the gauge variables $%
z^{\mu}(\tau, \sigma^r)$ \footnote{%
The multi-temporal approach is different by quantization methods like BRST,
in which one firstly quantizes all the variables, also the gauge ones, and
then makes the reduction to the physical ones at the quantum level by
selecting the states annihilated by a quantum version of the first-class
constraints (assuming that there is an ordering such that the quantum
constraints satisfy the same algebra as in the classical case, with the
quantum constraints located at the extreme right in the results of
commutators).}: they are considered as \textit{c-number generalized times}
in analogy to the treatment of time in the non-relativistic Schroedinger
equation, $i\, \hbar\, {\frac{{\partial}}{{\partial t}}}\, \psi(t, q) = \hat
H(q, \hat p)\, \psi(t, q)$. In this theory we have the c-number time $t$ and
the classical equality $E = H$ \footnote{$E$, the energy, is the generator
of the kinematical Poincar\'e group identified by the relativity principle,
while $H$ is the Hamiltonian governing the time evolution.} is realized with
$E \mapsto\, i\, \hbar\, {\frac{{\partial}}{{\partial t}}}$ and $H \mapsto\,
\hat H(q, \hat p)$. Therefore we send $\rho_{\mu}(\tau, \sigma^r) \mapsto
i\, \hbar\, {\frac{{\delta}}{{\delta\, z^{\mu}(\tau, \sigma^r)}}}$ and we
replace $\mathcal{G}_{\mu}$ with a suitably ordered self-adjoint operator ${%
\hat {\mathcal{G}}}_{\mu}$ depending upon the matter operators and the
c-number embeddings. Then the wave functional $\Psi(\tau; z^{\beta}(\tau,
\sigma^r) | \eta^r_i)$ must satisfy the following equations (as said in
Subsection A of the Section II the canonical Hamiltonian for the $\tau$%
-evolution is $M c$)

\begin{eqnarray}
i\, \hbar\, {\frac{{\partial}}{{\partial \tau}}}\, \Psi &=& \hat M c\, \Psi,
\nonumber \\
i\, \hbar\, {\frac{{\delta}}{{\delta\, z^{\mu}(\tau, \sigma^r)}}}\, \Psi &=&
{\hat {\mathcal{G}}}_{\mu}(\tau, \sigma^r)\, \Psi.  \label{5.1}
\end{eqnarray}

The theory is consistent if the quantum constraint operators ${\hat{\mathcal{%
H}}}_{\mu }(\tau ,\sigma ^{r})\,=\,i\,\hbar \,{\frac{{\delta }}{{\delta
\,z^{\mu }(\tau ,\sigma ^{r})}}}-{\hat{\mathcal{G}}}_{\mu }(\tau ,\sigma
^{r})$ are still Abelian like in the classical case, i.e. if we find an
ordering such that the integrability conditions for Eqs.(\ref{5.1}) $[{\hat{%
\mathcal{H}}}_{\mu }(\tau ,\sigma _{1}^{r}),{\hat{\mathcal{H}}}_{\nu }(\tau
,\sigma _{2}^{r})]=0$ hold \footnote{%
See Ref.\cite{10} for the modifications of the equations (\ref{5.1}) when
one has the more general integrable case $[{\hat{\mathcal{H}}}_{\mu }(\tau
,\sigma _{1}^{r}),{\hat{\mathcal{H}}}_{\nu }(\tau ,\sigma _{2}^{r})]=\int
d^{3}\sigma \,{\hat{C}}_{\mu \nu }^{\rho }(\sigma _{1}^{r},\sigma
_{2}^{r},\sigma ^{r})\,{\hat{\mathcal{H}}}_{\rho }(\tau ,\sigma ^{r})$}. The
other integrability condition for Eqs. (5.1) is $[\hat{M},{\hat{\mathcal{G}}}%
_{\mu }(\tau ,\sigma ^{r})]=0$.\medskip

In this functional Hilbert space one has to implement the second-class
constraints with the Gupta-Bleuler method. A non trivial problem is to find
the scalar product in the final physical Hilbert space. \medskip

The restriction of the solution of the coupled equations (\ref{5.1}) to the
surface $z^{\mu}(\tau, \sigma^r) = z^{\mu}_F(\tau, \sigma^r)$ of the
functional space of generalized times, with $z_F^{\mu}(\tau, \sigma^r)$ an
admissible 3+1 splitting of Minkowski space-time, gives the RQM in the
non-inertial frame classically defined by the gauge-fixings $z^{\mu}(\tau,
\sigma^r) - z^{\mu}_F(\tau, \sigma^r) \approx 0$ to the first-class
constraints $\mathcal{H}_{\mu}(\tau, \sigma^r) \approx 0$. The integrability
conditions imply that we can go from a non-inertial frame to a different one
preserving the quantum equivalence of the two descriptions. \medskip

When this program can be implemented, we have unitary equivalence of RQM in
every admissible non-inertial frame. In particular it should be possible to
implement the transformations (4.4) as time-dependent unitary
transformations connecting the rest-frame RQM to its non-inertial version.
Till now this has been achieved only for rotating coordinates and their
non-relativistic limit in Refs.\cite{10}.

\subsection{Alice and Bob Quantum Observers?}

As already said, the definition of a quantum dynamical observer is quite
problematic already at the non-relativistic level. At the relativistic level
the spatial non-separability implied by the second-class constraints
eliminating the internal center of mass inside the 3-space forces us to
include dynamical observers inside the isolated system as said in connection
with the Unruh-DeWitt detectors.\medskip

To avoid hybrid descriptions in which the trajectory of the detector is
classical but its interactions with the other objects are quantum, one has
to describe the detectors as macroscopic quantum many-body systems included
in the isolated quantum system. If one would be able to quantize these
systems, then the hybrid view would emerge due to notions like decoherence,
suggesting that the macroscopic quantum system has a quasi-classical
collective variable (the Pointer) following a semi-classical Newton-like
trajectory. See Ref.[23] for a discussion of the emergence of this classical
regime from the quantum one. \medskip

However, in the framework of non-relativistic quantum information theory the
problem of reference frames and of observers is very important \cite{24}
(see these papers and Ref. \cite{25} for the relativistic extension),
because for many tasks one needs information on clock synchronization, on
the alignment of distinct Cartesian axes and on the determination of global
positions. Connected problems are how two unrelated observers Alice and Bob
can compare measures of spin when they do not share a common reference
frame. In these cases the lack of a reference frame (or its level of
imprecision \cite{26}) is treated as a kind of decoherence (or quantum
noise) which can wash out all the quantum features of a measurement. See
Refs. \cite{24} for an attempt to define \textit{quantum reference frames}
by quantizing the measuring apparatus associated with a mathematical
observer (the previously quoted hybrid description).

\section{Final Remarks}

We have described the status of the theory of non-inertial frames in
Minkowski space-time developed by taking into account the problem of
relative times in relativistic bound states and the implications of Lorentz
signature for the relativistic collective variables. \medskip

In particular we have defined a new family of global non-inertial frames
with non-Euclidean instantaneous 3-spaces, which can be obtained from an
inertial frame by means of a point-dependent Lorentz transformation as
suggested by the locality principle.

\medskip

We have discussed properties of inertial and non-inertial either
mathematical or dynamical relativistic observers. \medskip

Already at the classical level we get a non-locality (and non-measurability)
of the canonical non-covariant relativistic external center of mass and a
spatial non-separability implied by the second-class constraints eliminating
the internal center of mass in the 3-spaces. At the quantum level this
non-locality and non-separability are at a deeper level with respect to the
standard discussion about the violation (Bell's inequalities) of the local
separable realism of Einstein in ordinary non-relativistic quantum mechanics.

\bigskip

The main open problems, besides the explicit construction of particle RQM in
non-inertial frames, are connected with the quantization of fields in
non-inertial frames:\medskip

A) find the rest-frame quantization of free scalar and transverse
electro-magnetic fields (see the second paper in Ref.\cite{14});\medskip

B) find the non-inertial frames in which the evolution of a massive scalar field
is unitary because the Bogoliubov transformation is of the Hilbert-Schmidt
type (solution of the Torre-Varadarajan no-go theorem \cite{27,28}) and try to
understand what happens to the notion of particle (see Ref.\cite{8}).

\appendix

\section{Calculations}

Given the parametrization of Eqs. (\ref{3.3})-(\ref{3.5}) of the Lorentz
transformation appearing in the embedding (\ref{3.1}), we must introduce new
quantities needed in the evaluation of the gradients of the embeddings and
then of the 4-metric of the non-inertial frame.

\bigskip

For the rotations we define the following quantities (in accord with the
notation of Subsection E of Section II)

\begin{eqnarray*}
{\tilde{\Omega}}_{(Ri)}(\tau ,\sigma ) &=&\Big({\tilde{R}}^{-1}\,{\frac{{%
\partial \,\tilde{R}}}{{\partial \,{\tilde{\alpha}}_{i}}}}\Big)(\tau ,\sigma
)=\left(
\begin{array}{ll}
0 & 0 \\
0 & \Omega _{(Ri)}=R^{-1}\,{\frac{{\partial \,R}}{{\partial \,{\tilde{\alpha}%
}_{i}}}}%
\end{array}%
\right) (\tau ,\sigma ), \\
&&{} \\
{\tilde{\Omega}}_{(Ri)}{}^{B}{}_{r}(\tau ,\sigma ) &=&\sum_{u}\,\delta
_{u}^{B}\,\Omega _{(Ri)\,ur}(\tau ,\sigma ),\qquad \Omega _{(Ri)uv}=-\Omega
_{(Ri)vu}{\buildrel {def}\over {=}}\epsilon _{uvw}\,\Omega _{(Ri)w}(\tau
,\sigma ), \\
&&{} \\
&&{} \\
&&\sum_{i}\,\alpha _{i}(\tau )\,\Omega _{(Ri)ur}(\tau ,\sigma )
{\buildrel {def}\over {=}}\,\sum_{v}\,\epsilon _{urv}\,{\check{\Omega}}%
_{(R)v}(\tau ,\sigma ), \\
&&{} \\
{\check{\Omega}}_{(R)v}(\tau ,\sigma ) &=&\sum_{i}\,\alpha _{i}(\tau
)\,\Omega _{(Ri)v}(\tau ,\sigma ) \\
&&{\check{\Omega}}_{(R)}(\tau ,\sigma )=\Big({\check{\Omega}}_{(R)v}(\tau
,\sigma )=\sum_{i}\,\alpha _{i}(\tau )\,\Omega _{(Ri)v}(\tau ,\sigma )\Big),
\end{eqnarray*}

\begin{eqnarray}
\Omega _{(r)}(\tau ,\sigma )_{uv} &=&(R^{-1}\,\partial _{r}\,R)_{uv}(\tau
,\sigma )=\sum_{i}\,\partial _{r}\,{\tilde{\alpha}}_{i}(\tau ,\sigma
)\,\Omega _{(Ri)uv}(\tau ,\sigma )=  \nonumber \\
&=&\partial _{r}\,f(\sigma )\,\sum_{i}\,\alpha _{i}(\tau )\,\Omega
_{(Ri)uv}(\tau ,\sigma )=\partial _{r}\,f(\sigma )\,\sum_{w}\,\epsilon
_{uvw}\,{\check{\Omega}}_{(R)w}(\tau ,\sigma ),  \nonumber \\
&&{}  \nonumber \\
&&{}  \nonumber \\
\Omega _{(\tau )}(\tau ,\sigma )_{uv} &=&(R^{-1}\,\partial _{\tau
}\,R)_{uv}(\tau ,\sigma )=f(\sigma )\,\sum_{i}\,{\dot{\alpha}}_{i}(\tau
)\,\Omega _{(Ri)uv}(\tau ,\sigma )=  \nonumber \\
&=&f(\sigma )\,\sum_{w}\,\epsilon _{uvw}\,{\frac{{\Omega _{(R)}(\tau ,\sigma
)}}{c}}\,{\hat{n}}^{w}(\tau ,\sigma ),  \label{a1}
\end{eqnarray}

\noindent where $\Omega^r(\tau, \sigma) = f(\sigma)\, \Omega_{(R)}(\tau,
\sigma)\, {\hat n}^r(\tau, \sigma)$ is the instantaneous angular velocity in
the point $(\tau, \sigma^r)$ (${\dot \alpha}_i(\tau) = {\frac{{d \alpha(\tau)%
}}{{d \tau}}} = {\frac{1}{c}}\, {\frac{{d\, \alpha_i(ct)}}{{dt}}}$)
\footnote{%
We have $0 < {\frac{{\Omega_{(R)}}}{c}} \leq 2\, max\, ({\dot \alpha}%
_1(\tau), {\dot \alpha}_2(\tau), {\dot \alpha}_3(\tau))$ as in Subsection E of Section II.} and the unit
3-vector $\hat n(\tau, \sigma)$ is the instantaneous rotation axis there.
\bigskip

For the Lorentz boosts we define the following derived quantities (we use ${%
\frac{{\partial\, \tilde \gamma}}{{\partial\, {\tilde \beta}^u}}} = {\tilde
\gamma}^3\, {\tilde \beta}^u$, ${\frac{{\partial\, \tilde \gamma\, {\tilde
\beta}^r}}{{\partial\, {\tilde \beta}^u}}} = \tilde \gamma\, (\delta^{ur} + {%
\tilde \gamma}^2\, {\tilde \beta}^u\, {\tilde \beta}^r)$, ${\frac{{\partial}%
}{{\partial\, {\tilde \beta}^u}}}\, {\frac{{{\tilde \gamma}^2}}{{\tilde
\gamma + 1}}} = {\tilde \gamma}^3\, {\frac{{\tilde \gamma - 1}}{{\tilde
\gamma + 1}}}\, {\tilde \beta}^u$, ${\tilde \gamma}^2\, {\vec {\tilde \beta}}%
^2 = {\tilde \gamma}^2 - 1$)

\begin{eqnarray*}
{\vec{\Omega}}_{(B)}(\tau ,\sigma )^{A}{}_{B} &=&\Big({\Omega} _{(B)u}(\tau
,\sigma )\Big)^{A}{}_{B} = \Big(\, \Big({\tilde{R}}^{-1}\,B^{-1}\,{\frac{{\
\partial \,B}}{{\partial \,{\tilde{\beta}}^{u}}}}\,\tilde{R}\Big) (\tau
,\sigma )\, \Big)^{A}{}_{B} =  \nonumber \\
&=&\Big(\, \Big({\tilde{R}}^{-1}\,{\tilde{\Omega}}_{(B)u}\,\tilde{R}\Big)%
(\tau ,\sigma )\, \Big)^{A}{}_{B},  \nonumber \\
&&{}  \nonumber \\
&&B^{-1} = \left(
\begin{array}{ll}
\tilde \gamma & - \tilde \gamma\, {\tilde \beta}^s \\
- \tilde \gamma\, {\tilde \beta}^r & \delta^{rs} + {\frac{{{\tilde \gamma}%
^2\, {\tilde \beta}^r\, {\tilde \beta}^s}}{{\tilde \gamma + 1}}}%
\end{array}
\right),  \nonumber \\
&&{}  \nonumber \\
&&{\frac{{\partial \,B}}{{\partial \,{\tilde \beta}^u }}}= \left(
\begin{array}{cc}
{\tilde \gamma}^3\, {\tilde \beta}^u & {\tilde \gamma}\, (\delta ^{tu} + {%
\tilde \gamma}^2\, {\tilde \beta}^u\, {\tilde \beta}^t) \\
\tilde \gamma\, (\delta ^{su} + {\tilde \gamma}^2\, {\tilde \beta}^u\, {%
\tilde \beta}^s) & {\frac{{\ {\tilde \gamma}^2}}{{\tilde \gamma + 1}}}\,
\left( \delta ^{su}\, {\tilde \beta}^t + {\tilde \beta}^s\, \delta^{tu} + {%
\frac{{\ {\tilde \gamma}^2\, (\tilde \gamma + 2)\, {\tilde \beta}^s\, {%
\tilde \beta}^t\, {\tilde \beta}^u }}{{\tilde \gamma + 1}}} \right)%
\end{array}
\right)
\end{eqnarray*}

\begin{eqnarray*}
{\tilde{\Omega}}_{(B)u} &=&B^{-1}\,{\frac{{\partial \,B}}{{\partial \,{%
\tilde{\beta}}^{u}}}}=\left(
\begin{array}{l}
{0} \\
\tilde{\gamma}\,\Big(\delta ^{um}+{\frac{{\tilde{\gamma}}^{2}} {{\tilde{%
\gamma}+1}}}\,{\tilde{\beta}}^{u}\,{\tilde{\beta}}^{m}\Big)%
\end{array}
\right.  \nonumber \\
&&{}  \nonumber \\
&&\left.
\begin{array}{l}
\tilde \gamma\, \Big(\delta^{us} + {\frac{{{\tilde \gamma}2}}{{\tilde \gamma
+ 1}}} \tilde{\beta}^{u}{\tilde{\beta}}^{s}\Big) \\
{\frac{ {{\tilde \gamma}^2}}{{\ \tilde \gamma + 1 }}} \,\Big(\delta^{um}\, {%
\tilde \beta}^s - \delta ^{us}\,{\tilde{\beta}}^m\Big)%
\end{array}
\right) ,  \nonumber \\
&&{}  \nonumber \\
&&{}  \nonumber \\
\Omega _{(B)u} &=& {\tilde R}^{-1}\, {\tilde \Omega}_{(B)u}\, \tilde R =
\left(
\begin{array}{l}
{0} \\
\tilde{\gamma}\,R_{vm}^{T}\,\Big(\delta ^{um}+{\frac{{{\tilde{\gamma}} ^{2}}%
}{{\tilde{\gamma}+1}}}\,{\tilde{\beta}}^{u}\,{\ \tilde{\beta}}^{m}\Big)%
\end{array}
\right.  \nonumber \\
&&{}  \nonumber \\
&&\left.
\begin{array}{l}
\tilde{\gamma}\,\Big(\,\delta ^{us}+{\frac{\tilde{\gamma}^2}{{\tilde{\gamma}%
+1}}}\,{\tilde{\beta}}^{u}\,{\tilde{\beta}}^{s}\Big)\,R_{sn} \\
{\frac{ {{\tilde \gamma}^2}}{{\ \tilde \gamma + 1 }}}\, R_{vm}^{T}\,\Big(%
\delta ^{um}\,{\tilde{\beta}}^s - \delta ^{us}\,{\tilde{\beta}}^m \Big)%
\,R_{sn}%
\end{array}
\right) ,  \nonumber \\
\end{eqnarray*}

\begin{eqnarray}
{\vec{\tilde{\beta}}}\cdot {\vec{\Omega}}_{(B)} &=& \left (
\begin{array}{l}
{0} \\
{\tilde \gamma}^2\, R_{vm}^{T}\, \tilde{\beta}^{m}%
\end{array}
\right. \left.
\begin{array}{l}
{\tilde \gamma}^2\, {\tilde{\beta}}^{s}\,R_{sn} \\
0%
\end{array}
\right) ,  \nonumber \\
&&{}  \nonumber \\
{\dot{\vec{\tilde{\beta}}}}\cdot {\vec{\Omega}}_{(B)} &=&\left(
\begin{array}{l}
{0} \\
\tilde{\gamma}\,R_{vm}^{T}\,\Big({\dot{\tilde{\beta}}}{}^{m}+ {\frac{{\ {%
\tilde \gamma}^2 }}{{\tilde \gamma + 1}}}\, {\dot{\vec{ \tilde{\beta}}}}%
\cdot {\vec{\tilde{\beta}}}\,{\tilde{\beta}}^{m}\Big)%
\end{array}
\right.  \nonumber \\
&&{}  \nonumber \\
&&\left.
\begin{array}{l}
\tilde{\gamma}\,\Big(\,{\dot{\tilde{\beta}}}{}^{s}+{\frac{\tilde{\gamma}^2}{{%
\tilde{\gamma}+1}}}\,{\dot{\vec{\tilde{\beta}}}}\cdot {\vec{\tilde{\beta}}}\,%
{\tilde{\beta}}^{s}\Big)\,R_{sn} \\
\, {\frac{{\ {{\tilde{\gamma}}^{2}} }}{{\tilde \gamma + 1 }}}\, R_{vm}^{T}\, %
\Big( {\dot{\tilde{\beta}}}{}^m\, {\tilde \beta}^s - {\dot{\tilde{\beta}}}%
{}^s\, {\tilde \beta}^m\Big)\,R_{sn}%
\end{array}%
\right).  \label{a2}
\end{eqnarray}

\bigskip

As a consequence of the previous notations, we get the following expression
for the gradients of the Lorentz matrix appearing in Eq.(3.1) (${\hat \sigma}%
^s = \sigma^s/\sigma$, $f^{^{\prime }}(\sigma) = {\frac{{d f(\sigma)}}{{d
\sigma}}}$)

\begin{eqnarray*}
\partial_{\tau}\, \Lambda^A{}_r(\tau, \sigma) &=& \Lambda^A{}_B(\tau,
\sigma)\, \Big(\Lambda^{-1}\, \partial_{\tau}\, \Lambda\Big)^B{}_r(\tau,
\sigma) =  \nonumber \\
&=&\Lambda^A{}_B(\tau, \sigma)\, \Big[{\tilde R}^{-1}\, B^{-1}\, \Big(%
f(\sigma)\, B\, \sum_i\, {\dot \alpha}_i(\tau)\, {\frac{{\partial\, \tilde R}%
}{{\partial\, {\tilde \alpha}_i}}} +  \nonumber \\
&+& g(\sigma)\, {\dot {\vec \beta}}(\tau) \cdot {\frac{{\partial\, B}}{{%
\partial\, {\vec {\tilde \beta}}}}}\, \tilde R\Big)\Big]^B{}_r(\tau, \sigma)
=  \nonumber \\
&=&\Lambda^A{}_B(\tau, \sigma)\, \Big[f(\sigma)\, \sum_i\, {\dot \alpha}%
_i(\tau)\, {\tilde \Omega}_{(Ri)}{}^B{}_r + g(\sigma)\, {\dot {\vec \beta}}%
(\tau) \cdot {\vec \Omega}_{(B)}{}^B{}_r\Big](\tau, \sigma) =  \nonumber \\
&=&\Lambda^A{}_B(\tau, \sigma)\, \Big[ f(\sigma)\, {\frac{{\Omega_{(R)}}}{c}}%
\, \sum_{uv}\, \delta^B_u\, \epsilon_{urv}\, {\hat n}^v + g(\sigma)\, {\dot {%
\vec \beta}}(\tau) \cdot {\vec \Omega}_{(B)}{}^B{}_r \Big](\tau, \sigma),
\nonumber \\
\end{eqnarray*}

\begin{eqnarray}
\partial_s\, \Lambda^A{}_r(\tau, \sigma) &=&\Lambda^A{}_B(\tau, \sigma)\, %
\Big(\Lambda^{-1}\, \partial_s\, \Lambda\Big)(\tau, \sigma){}^B{}_r =
\nonumber \\
&=&\Lambda^A{}_B(\tau, \sigma)\, {\hat \sigma}^s\, \Big[f^{^{\prime
}}(\sigma)\, \sum_i\, \alpha_i(\tau)\, \Big({\tilde R}^{-1}\, {\frac{{%
\partial\, \tilde R}}{{\partial\, {\tilde \alpha}_i}}}\Big) +  \nonumber \\
&+& g^{^{\prime }}(\sigma)\, {\vec \beta}(\tau) \cdot \Big({\tilde R}^{-1}\,
B^{-1}\, {\frac{{\partial\, B}}{{\partial\, {\vec {\tilde \beta}}}}}\,
\tilde R\Big)\Big](\tau, \sigma){}^B{}_r =  \nonumber \\
&=&\Lambda^A{}_B(\tau, \sigma)\, {\hat \sigma}^s\, \Big[f^{^{\prime
}}(\sigma)\, \sum_i\, \alpha_i(\tau)\, {\tilde \Omega}_{(Ri)}{}^B{}_r +
g^{^{\prime }}(\sigma)\, \vec \beta(\tau) \cdot {\vec \Omega}_{(B)}{}^B{}_r%
\Big](\tau, \sigma) =  \nonumber \\
&=&\Lambda^A{}_B(\tau, \sigma)\, {\hat \sigma}^s\, \Big[f^{^{\prime
}}(\sigma)\, \sum_{uv}\, \delta^B_u\, \epsilon_{urv}\, {\check \Omega}_{(R)
v} + g^{^{\prime }}(\sigma)\, \vec \beta(\tau) \cdot {\vec \Omega}%
_{(B)}{}^B{}_r\Big](\tau, \sigma).  \label{a3}
\end{eqnarray}

\bigskip

Therefore we get the following expression for the gradients of the embedding
(\ref{3.1})

\begin{eqnarray*}
z^{\mu}_{\tau}(\tau, \sigma^u) &=& {\dot x}^{\mu}(\tau) + \epsilon^{\mu}_A\,
\partial_{\tau}\, \Lambda^A{}_r(\tau, \sigma)\, \sigma^r =  \nonumber \\
&=&\epsilon^{\mu}_A\, \Big({\dot f}^A(\tau) + \sigma\, \Lambda^A{}_B\, \Big[%
f(\sigma)\, {\frac{{\Omega_{(R)}}}{c}}\, \sum_u\, \delta^B_u\, (\hat \sigma
\times \hat n)^u +  \nonumber \\
&+& g(\sigma)\, {\dot {\vec \beta}}(\tau) \cdot {\vec \Omega}%
_{(B)}{}^B{}_r\, {\hat \sigma}^r\Big]\Big)(\tau, \sigma) =  \nonumber \\
&{\buildrel {def}\over {=}}& \epsilon^{\mu}_A\, \Lambda^A{}_B(\tau,
\sigma)\, \Big((\Lambda^{-1})^B{}_C\, {\dot f}^C(\tau) + F^B\Big)(\tau,
\sigma),  \nonumber \\
\end{eqnarray*}

\begin{eqnarray}
z^{\mu}_r(\tau, \sigma^u) &=& \epsilon^{\mu}_A\, \Big(\Lambda^A{}_r(\tau,
\sigma) + \partial_r\, \Lambda^A{}s(\tau, \sigma)\, \sigma^s\Big) =
\nonumber \\
&=&\epsilon^{\mu}_A\, \Lambda^A{}_B(\tau, \sigma)\, \Big(\delta^B_r +
\sigma\, {\hat \sigma}^r\, \Big[f^{^{\prime }}(\sigma)\, \sum_u\,
\delta^B_u\, (\hat \sigma \times {\check \Omega}_{(R)})^u +  \nonumber \\
&+& g^{^{\prime }}(\sigma)\, \vec \beta(\tau) \cdot {\vec \Omega}%
_{(B)}{}^B{}_n\, {\hat \sigma}^n\Big]\Big)(\tau, \sigma) =  \nonumber \\
&{\buildrel {def}\over {=}}& \epsilon^{\mu}_A\, \Lambda^A{}_B(\tau,
\sigma)\, G^B_r(\tau, \sigma),  \label{a4}
\end{eqnarray}

\noindent with the following identifications

\begin{eqnarray*}
&&\Big(\Lambda^{-1}(\tau, \sigma)\Big)^{\tau}{}_H\, {\dot f}^H(\tau) +
F^{\tau}(\tau, \sigma) =  \nonumber \\
&&= \Big(\alpha(\tau)\, \gamma_x(\tau)\, \tilde \gamma\, [1 - {\vec {\tilde
\beta}} \cdot {\vec \beta}_x(\tau)] + \sigma\, g(\sigma)\, {\dot {\vec \beta}%
}(\tau) \cdot {\vec \Omega}_{(B)}{}^{\tau}{}_n\, {\hat \sigma}^n\Big)%
(\tau,\sigma),  \nonumber \\
&&{}  \nonumber \\
&&\Big(\Lambda^{-1}(\tau, \sigma)\Big)^r{}_H\, {\dot f}^H(\tau) + F^r(\tau,
\sigma) =  \nonumber \\
&&= \Big(\alpha(\tau)\, \gamma_x(\tau)\, \sum_v\, R^T_{rv}\, \Big[%
\beta_x^v(\tau) - \tilde \gamma\, {\tilde \beta}^v\, \Big(1 - {\frac{{\tilde
\gamma\, {\vec {\tilde \beta}} \cdot {\vec \beta}_x(\tau)}}{{\tilde \gamma +
1}}}\Big)\Big] +  \nonumber \\
&&+ \sigma\, \Big[f(\sigma)\, {\frac{{\Omega_{(R)}}}{c}}\, (\hat \sigma
\times \hat n)^r + g(\sigma)\, {\dot {\vec \beta}}(\tau) \cdot {\vec \Omega}%
_{(B)}{}^r{}_n\, {\hat \sigma}^n \Big]\Big)(\tau, \sigma),  \nonumber \\
\end{eqnarray*}

\begin{eqnarray}
&&G^s_r(\tau, \sigma) = \delta^s_r + \sigma\, {\hat \sigma}^r\, \Big[%
f^{^{\prime }}(\sigma)\, (\hat \sigma \times {\check \Omega}_{(R)})^s +
g^{^{\prime }}(\sigma)\, \vec \beta(\tau) \cdot {\vec \Omega}%
_{(B)}{}^s{}_n\, {\hat \sigma}^n\Big](\tau, \sigma),  \nonumber \\
{}&&  \nonumber \\
&&G^{\tau}_r(\tau, \sigma) = \sigma\, {\hat \sigma}^r\, g^{^{\prime
}}(\sigma)\, \vec \beta(\tau) \cdot {\vec \Omega}_{(B)}(\tau,
\sigma){}^{\tau}{}_n\, {\hat \sigma}^n.  \label{a5}
\end{eqnarray}

\bigskip For the evaluation of the 4-metric we also need

\begin{eqnarray}
\Lambda^{\tau}{}_B(\tau, \sigma) &-& \sum_r\, \beta^r_x(\tau)\,
\Lambda^r{}_B(\tau, \sigma) = \Big(\tilde \gamma\, (1 - {\vec \beta}_x(\tau)
\cdot {\vec {\tilde \beta}})\, \delta^{\tau}_B +  \nonumber \\
&+& \sum_{uv}\, \Big[\tilde \gamma\, (1 - {\frac{{\tilde \gamma}}{{\tilde
\gamma + 1}}}\, {\vec \beta}_x(\tau) \cdot {\vec {\tilde \beta}})\, {\tilde
\beta}^u - \beta^u_x(\tau)\Big]\, R_{uv}\, \delta^v_B\Big)(\tau, \sigma).
\label{a6}
\end{eqnarray}

\bigskip

The unit normal $l^{\mu}(\tau, \sigma^u)$ to the instantaneous 3-space $%
\Sigma_{\tau}$ is

\begin{eqnarray}
l^{\mu}(\tau, \sigma^u) &=& \epsilon^{\mu}_D\, l^D(\tau, \sigma^u) = \Big({%
\frac{1}{\sqrt{h}}}\, \epsilon^{\mu}{}_{\alpha\beta\gamma}\, z^{\alpha}_1\,
z^{\beta}_2\, z^{\gamma}_3\Big)(\tau, \sigma^u) =  \nonumber \\
&=&\Big({\frac{1}{\sqrt{h}}}\, \epsilon^{\mu}{}_{\alpha\beta\gamma}\,
\epsilon^{\alpha}_A\, \epsilon^{\beta}_B\, \epsilon^{\gamma}_C\,
\Lambda^A{}_E\, \Lambda^B{}_F\, \Lambda^C{}_G\, G^E_1\, G^F_2\, G^G_3\Big)%
(\tau, \sigma^u).  \label{a7}
\end{eqnarray}

\bigskip

Using Eq.(\ref{a7}) for the unit normal to the 3-space, the lapse function $%
N = 1 + n$ defined in Subsection A of Section II turns out to be (we have $%
\epsilon_{\mu\alpha\beta\gamma}\, \epsilon^{\mu}_D\, \epsilon^{\alpha}_A\,
\epsilon^{\beta}_B\, \epsilon^{\gamma}_C = \epsilon_{DABC}$ and $%
\epsilon_{DABC}\, \Lambda^D{}_W\, \Lambda^A{}_E\, \Lambda^B{}_F\,
\Lambda^C{}_G = \epsilon_{WEFG}$)

\begin{eqnarray}
1 + n(\tau, \sigma^u) &=&\epsilon\, \Big(l^{\mu}\, \eta_{\mu\nu}\,
z^{\nu}_{\tau}\Big)(\tau, \sigma^u) =  \nonumber \\
&=&\Big({\frac{1}{\sqrt{h}}}\, \epsilon_{\mu\alpha\beta\gamma}\,
\epsilon^{\mu}_D\, \epsilon^{\alpha}_A\, \epsilon^{\beta}_B\,
\epsilon^{\gamma}_C\, \Lambda^D{}_W\, \Lambda^A{}_E\, \Lambda^B{}_F\,
\Lambda^C{}_G  \nonumber \\
&&\Big[(\Lambda^{-1})^W{}_H\, {\dot f}^H(\tau) + F^W\Big]\, G^E_1\, G^F_2\,
G^G_3 \Big)(\tau, \sigma^u) =  \nonumber \\
&=& \Big({\frac{1}{\sqrt{h}}}\, \epsilon_{WEFG}\, \Big[(\Lambda^{-1})^W{}_H%
\, {\dot f}^H(\tau) + F^W\Big]\, G^E_1\, G^F_2\, G^G_3 \Big)(\tau, \sigma^u).
\label{a8}
\end{eqnarray}

\vfill\eject

\end{document}